\documentclass[useAMS,usenatbib,pdflscape]{mn2e}
\usepackage{graphicx}
\usepackage{epstopdf}
\usepackage{amsmath}
\usepackage{amssymb}
\usepackage{rotating}
\usepackage{pdflscape}
\usepackage{appendix}

\UseRawInputEncoding

\title[]{Dynamical and thermal properties of the parsec-scale gases spherically accreted onto low luminous active galactic nuclei}
\author[Sun \& Yang]{Han-Wen Sun \& Xiao-Hong Yang\thanks{Corresponding author: Xiao-Hong Yang}\\
 Department of Physics, Chongqing University, Chongqing 400044, China; yangxh@cqu.edu.cn\\
}

\begin{document}

\pagerange{\pageref{firstpage}--\pageref{lastpage}} \pubyear{20**}

\maketitle

\label{firstpage}
\def\LSUN{\rm L_{\odot}}
\def\MSUN{\rm M_{\odot}}
\def\RSUN{\rm R_{\odot}}
\def\MSUNYR{\rm M_{\odot}\,yr^{-1}}
\def\MSUNS{\rm M_{\odot}\,s^{-1}}
\def\MDOT{\dot{M}}

\begin{abstract}
We analytically study the dynamical and thermal properties of the optically-thin gases at the parsec-scale when they are spherically accreted onto low luminous active galactic nuclei (LLAGNs). The falling gases are irradiated by the central X-ray radiation with the Compton temperature of 5--15$\times10^7$ K. The radiative heating/cooling and the bulge stellar potential in galaxies are taken into account. We analyze the effect of accretion rate, luminosity, gas temperature, and Compton temperature on steady solutions of dynamical and thermal properties. The steady solutions are obviously different from Bondi solution. Compared to our models, the Bondi model underestimates the accretion rate. We give the boundary between thermal stability and instability. The boundary is significantly affected by Compton temperature. When Compton temperature is higher, the falling gases tend to become thermally unstable. When thermal instability takes place in the irradiated gases, the gases become two phases (i.e. hot gases and cool gases) and the hot gases may become outflows. This effect may reduce the accretion rates.
\end{abstract}


\section{INTRODUCTION}
For decades, the Bondi model has been the most basic tool to describe the spherical and steady accretion onto a point mass source (Bondi 1952), such as a supermassive black hole (SMBH) at the center of galaxies. Because of its inherent simplicity, the Bondi model is often useful for us to understand the basic physical nature of the accretion phenomenon. When density and temperature at infinity are given, the mass accretion rate on the point source is obtained based on Bondi model. Therefore, the classical Bondi model is often used to estimate the mass accretion rate on the SMBH in the following two cases. (1) In observations, the mass accretion rate of SMBH is often estimated by using the observed values of gas density and temperature in the vicinity of the SMBH. (2) In numerical simulations of studying galaxy formation and cosmos evolution, physical processes around the SMBH can not be solved, and therefore Bondi model is used to estimate the mass accretion rate (e.g. Loewenstein et al. 2001; Baganoff et al. 2003; Pellegrini 2005,2010; Di Matteo et al. 2005,2008; Allen et al. 2006; Barai et al. 2011; McNamara et al. 2011; Wong et al. 2014; Russell et al. 2015; Beckmann et al. 2018). The mass accretion rate is an important parameter to estimate the SMBH luminosity. This is very important for the study of the feedback effect of active galactic nuclei (AGNs) (e.g. Yu \& Tremaine 2002; Kurosawa \& Proga 2009; Novak et al. 2011; Xie \& Yuan 2012; Gan et al. 2014).

However, for realistic accreting processes, the Bondi model may be too simple and lacks many necessary details. For example, the following factors make the classical Bondi model not applicable to accurately estimate the mass accretion rate. (1) When the falling gas slowly rotates, the mass accretion rate can be significantly reduced compared to the accretion of the Bondi model (Proga \& Begelman 2003). (2) The radiation feedback from the accreting flow around SMBHs can reduce the mass accretion rate (e.g. Ciotti \& Ostriker 2007; Yang \& Bu 2018a; Bu \& Yang 2018). (3) Stars and dark matter in a galaxy can affect the dynamics of spherical accretion (Korol et al. 2016; Ciotti \& Pellegrini 2017,2018; Ciotti \& Ziaee Lorazd 2018; Samadi et al. 2019). (4) Thermal instability may occur in the parsec-scale gas irradiated by the accretion disk around SMBHs and then influences the dynamics and thermodynamics of spherical accretion (e.g. Ostriker et al. 1976; Cowie et al. 1978; Krolik \& London 1983; Moscibrodzka \& Proga 2013; Waters \& Proga 2019; Dannen et al. 2020; Bu, Yang \& Zhu 2020). The irradiated parsec-scale gas by AGNs could become into a two-phase/cold-hot accretion flow due to thermal instability (Moscibrodzka \& Proga 2013; Bu, yang \& Zhu 2020), which is interesting to understand the material feeding of AGNs. Thermal instability may be important not only in the parsec-scale (e.g. Moscibrodzka \& Proga 2013; Bu, Yang \& Zhu 2020) but also in the kpc-scale or the intracluster/circumgalactic medium (e.g. McCourt et al 2012; Choudhury \& Sharma 2015; Sobacchi \& Sormani 2019; Choudhury et al. 2019; Das et al. 2021). The local thermal instability could trigger cold gas to form in the intracluster/circumgalactic medium (e.g. Choudhury et al. 2019). The cold gas is crucial to understand the formation of galaxies.

The accretion flow onto a SMBH has two modes: the cold accretion flows (such as the standard thin disk model and the slim disk model) and the hot accretion flows. The standard thin disk is often used to describe quasars or the soft state of X-ray binaries, whose luminosity and accretion rate are relatively high (Shakura \& Sunyaev 1973). Quasar can emit a great number of photons in the optical--\textit{UV} bands and a small number of X-ray photons (e.g. Proga et al. 2000; Proga 2007; Kurosawa \& Proga 2009), whose Compton temperature is about $\sim2.7\times10^7$ K (i.e. 10 kev). The hot accretion flows are often used to describe low luminous active galactic nuclei (LLAGNs) or the hard state of X-ray binaries, whose luminosity and accretion rate are relatively low (Narayan \& Yi 1994; Yuan \& Narayan 2014). LLAGNs can emit a great number of high-energy photons, whose Compton temperature is about 5--15$\times10^7$ K (Xie et al. 2017). In this paper, the focus of our study is the parsec-scale gas accreted on LLAGNs.

When the gas at the parsec-scale is irradiated by X-ray photons from AGNs, thermal instability may take place (Field 1965; Krolik \& London 1983; Barai et al 2011; Mo\'{s}cibrodzka \& Proga 2013). Krolik \& London (1983) studied the thermal instability of spherical accretion to a quasar and Mo\'{s}cibrodzka \& Proga (2013) implemented numerical simulations. Mo\'{s}cibrodzka \& Proga (2013) have observed that the gas is thermally and convectively unstable within the region of 0.1pc--200pc. In their simulations, Compton temperature is set to be $\sim2.7\times10^7$ k, which is applicable to a quasar. However, for a quasar, majority of the emitted photons are \textit{UV} photons, whose force effects are not be neglected in the real case (e.g. Proga 2007; Proga et al. 2008; Kurosawa \& Proga 2009; Ram\'{i}rez-Vel\'{a}sques et al. 2019). When Compton scattering force of \textit{UV} photons is included, Bu, Yang \& Zhu(2020) have also observed thermal instability from numerical simulations.

In this paper, we extend Krolik \& London's work (1983) to spherical accretion on LLAGNs. Because LLAGNs have higher Compton temperature than quasars, we analyze the dynamical and thermodynamic properties of parsec-scale spherical accretion in the case with higher Compton temperature and estimate the effect of Compton temperature on thermal instability. Our results are applicable to estimate the mass accretion rate of LLAGNs in observations and in the numerical simulations of galaxy formation and evolution.

The paper is organized as follows. In section 2, we describe our model and method. In section 3, we present our results and related discussions. In section 4, we give a summary and discussions.

\section{MODEL AND METHOD}
\subsection{Basic model and equations}
In this paper, we analytically study how the parsec-scale gas is accreted onto LLAGNs, which can be described by a hot accretion flow (Narayan \& Yi 1994; Yuan \& Narayan 2014). In this case, The parsec-scale gas is irradiated by the X-ray photons from LLAGNs, and so the radiative heating and cooling are considered. Besides, the bulge stellar potential is also included. We also assume that the parsec-scale gas has low angular momentum. Therefore, the effect of angular momentum can be neglected at the parsec-scale. In order to simplify our model, the accretion flow is set to be spherically symmetric and independent on time. With these assumptions, we can get the time-independent hydrodynamic equations. The continuity equation of mass gives
\begin{equation}
\frac{{\rm{d}}}{{{\rm{d}}r}}\left( {4\pi {r^2}\rho v} \right) = 0,
\end{equation}
where $r$, $\rho$ and $v$ are the radius, the density and the velocity of the accreting gas, respectively. This equation implies that the mass accretion rate $\dot M = 4\pi {r^2}\rho v$ is constant with the radius. The momentum equation gives
\begin{equation}
v\frac{{{\rm{d}}v}}{{{\rm{d}}r}} =  - \frac{1}{\rho }\frac{{{\rm{d}}p}}{{{\rm{d}}r}} - g,
\end{equation}
where $p$ is the gas pressure and $g$ is the sum of gravity and radiation pressure force exerted on unit mass. Previous studies have found that the gravitational effect of bulge stars should not be neglected at the parsec-scale, such as beyond 1 parsec (Bu et al. 2016; Yang \& Bu 2018b). Since our computational region covers the sub-parsec and parsec scale, the gravitational force of stars in bulge is included in the total gravity. The total gravitational potential is given by $\psi_{\rm t}=\psi _{\rm BH}+\psi _{\rm{S}}$, where $\psi _{\rm BH}$ is the Paczy\'{n}ski-Wiita potential (${\psi _{{\rm{BH}}}} = {\rm{ - }}G{M_{{\rm{BH}}}}/(r - {r_{\rm{s}}})$) of a black hole(Paczy\'{n}ski \& Wiita 1980) and $\psi _{\rm{S}}$ is the bulge stellar potential. According the $M_{\rm BH}$--$\sigma$ relation ($M_{\rm BH}$ is the black hole mass and $\sigma$ is the dispersion velocity of stars, respectively) (Greene \& Ho 2006), Bu et al. (2016) gave ${\psi _{\rm{S}}}{\rm{ = }}{\sigma ^2}\ln r + C$, where $C$ is a constant. We set $\sigma$ to be 200 km/s for ${M_{{\rm{BH}}}}{\rm{ = }}{10^8}{M_ \odot }$ (Greene \& Ho 2006). Therefore, $g$ is given by
\begin{equation}
g = d\psi /dr - {f_{\rm{X}}}\frac{{G{M_{{\rm{BH}}}}}}{{{r^2}}} = \left[ {\frac{{{r^2}}}{{{{(r - {r_{\rm{s}}})}^2}}} - {f_{\rm{X}}}} \right]\frac{{G{M_{{\rm{BH}}}}}}{{{r^2}}} + \frac{{{\sigma ^2}}}{r},
\end{equation}
where $f_{\rm X}$ is the Eddington ratio of the X-ray luminosity ($L_{\rm X}$) of LLAGNs and $r_{\rm s}$ is the Schwarzschild radius. Since we focus on LLAGNs in this study, the value of $f_{\rm X}$ is always below 0.02, which means that the radiation pressure is not important here. The effect of the radiation pressure on the solutions is negligible.

Strong X-ray radiation can be produced near SMBHs and the Compton temperature ($T_{\rm C}$) of LLAGNs is about $\sim$(5--15)$\times10^7{\rm{ K}}$ (Xie et al. 2017). The radiation thermally influences the properties of the gas at the parsec scale. The energy equation is written as
\begin{equation}
\rho v\frac{{\rm{d}}}{{{\rm{d}}r}}(\frac{e}{\rho }) - v\frac{p}{\rho }\frac{{{\rm{d}}\rho }}{{{\rm{d}}r}} = {n^2}S,
\end{equation}
where $e = p/\left( {\gamma  - 1} \right)$ is the energy density of gas with the adiabatic index of $\gamma=5/3$,  $n$ is the number density of the gas, and $n^2S$ is the net heating rate by the radiation heating and cooling, respectively. In addition, we also adopt the equation of state, such as $p = \rho {k_{\rm{B}}}T/\mu {m_{\rm{p}}} = n{k_{\rm{B}}}T$, where $\mu$, $k_{\rm B}$, $m_{\rm p}$, and $T$ are the mean molecular weight, the Boltzmann constant, the proton mass and the gas temperature, respectively. We set $\mu=0.61$ in this paper.

The net heating rate is expressed as
\begin{equation}
n^2S = n^2({\Gamma _{\rm{C}}} + {\Gamma _{\rm{X}}} - {\Lambda _{\rm{B}}} - {\Lambda _{\rm{L}}})\rm{ }({\rm{ergs}} \cdot {\rm{c}}{{\rm{m}}^{ - 3}} \cdot {\rm{ }}{{\rm{s}}^{ - 1}}),
\end{equation}
where ${n^2}{\Gamma _{\rm{C}}}$ and ${n^2}{\Lambda _{\rm{B}}}$ are the rate of Compton heating/cooling and bremsstrahlung cooling, respectively. They respectively read
\begin{equation}
{\Gamma _{\rm{C}}} = 3.56 \times {10^{ - 35}}\xi ({T_{\rm{C}}} - T)
\end{equation}
and
\begin{equation}
{\Lambda_{\rm B}} = 3.3 \times {10^{ - 27}}{T^{0.5}},
\end{equation}
where $\xi$ (=$L_{\rm X}/(nr^2)$) is the ionization parameter of gas. ${n^2}{\Gamma _{\rm{X}}}$ is the net rate of X-ray photoionization heating and recombination cooling and ${n^2}{\Lambda _{\rm{L}}}$ is the rate of line cooling. In order to get the analytical formulae of $\Gamma_{\rm X}$ and $\Lambda _{\rm L}$, Blondin (1994) used the photoionization code to estimate the heating/cooling of an optically-thin gas with cosmic abundance and gave their analytical formulae of $\Gamma _{\rm X}$ and $\Lambda _{\rm L}$, during which a 10 keV bremsstrahlung spectrum is used to illuminate the gas. These formulae are in $25\% $ agreement with numerical simulations. For the sake of convenience, Blondin's formulae are copied here. $\Gamma _{\rm X}$ and $\Lambda _{\rm L}$ respectively read
\begin{equation}
{\Gamma _{\rm{X}}} = 1.5 \times {10^{ - 21}}{\xi ^{0.25}}{T^{ - 0.5}}(1 - \frac{T}{{4{T_{\rm{C}}}}})
\end{equation}
and
\begin{equation}
{\Lambda _{\rm{L}}} = 1.7 \times {10^{ - 18}}\exp ( - \frac{{1.3 \times {{10}^5}}}{T}){\xi ^{ - 1}}{T^{ - 0.5}} + {10^{ - 24}}.
\end{equation}
For LLAGNs, the Compton temperature of X-ray radiation is higher than 10 keV. However, the X-ray photoionization heating is not important in most of our models. Therefore, we still adopt the above formulae to calculate the X-ray photoionization heating.

\subsection{Two subclasses of models and model setup}
We set the black hole mass to be $M_{\rm BH} = 10^8 M_{\odot}$. Table 1 lists the parameters of our models. In Table 1, columns (1)--(6) are model name, the Compton temperature of radiation, the gas temperature at the outer boundary, the accretion rate, the LLAGNs luminosity, and the stellar velocity dispersion $\sigma$, respectively.

The models in Table 1 are classified into two subclasses, i.e. models A1--A8 and models B1--B9. Here, models A1--A8 are called A-type models while models B1--B9 are called B-type models. The radial range of our calculation is $200{r_{\rm{s}}}$ $\le r \le 1.44 \times {10^6}{r_{\rm{s}}}$ for the A-type models while $2000{r_{\rm{s}}}$ $\le r \le 1.44 \times {10^6}{r_{\rm{s}}}$ for the B-type models. In all of our models, the gas at the parsec-scale is assumed to have low angular momentum. Here, we use the ``circularization'' radius ($r_{\rm cir}$) to describe the angular momentum of gas, i.e. the angular momentum of gas equals the angular momentum of Kepler rotation at $r_{\rm cir}$. Inside the circularization radius, magnetorotational instability (MRI) can effectively work and the MRI-driven angular momentum transfer makes the gas continuously fall onto the black hole (e.g. Stone \& Pringle 2001). We set the circularization radius to be 50$r_{\rm s}$, which is much less than the inner boundary of the computational domain. Then, the effect of angular momentum is neglected within the computational region.
\begin{table}
\begin{center}

\caption[]{Summary of Models}

\begin{tabular}{ccccccccccc}
\hline\noalign{\smallskip} \hline\noalign{\smallskip}

 Model ID&  ${T_{\rm{C}}}$ & $T (r_{\rm{out}})$  &  $\dot{M}$ & $L_{\rm{X}}$ &$\sigma $ \\
    &   (K)   & (K) & (${\dot M_{Edd}}$) & (${L_{Edd}}$)& (km/s) \\
(1) & (2)     & (3) &  (4)      &     (5) & {6}   \\

\hline\noalign{\smallskip}
A1   &$5 \times {10^7}$ &$1 \times {10^7}$&0.013 &0.006 &200  \\
A2   &$5 \times {10^7}$ &$1 \times {10^7}$&0.022 &0.01  &200  \\
A3   &$5 \times {10^7}$ &$1 \times {10^7}$&0.045 &0.02  &200  \\
A4   &$5 \times {10^7}$ &$5 \times {10^6}$&0.022 &0.01  &200  \\
A5   &$5 \times {10^7}$ &$2 \times {10^7}$&0.022 &0.01  &200  \\
A6   &$1 \times {10^8}$ &$1 \times {10^7}$&0.013 &0.006 &200  \\
A7   &$1.5 \times {10^8}$&$1 \times {10^7}$&0.013 &0.006&200  \\
A8   &$5 \times {10^7}$ &$1 \times {10^7}$&0.022 &0.01  &0    \\
\hline\noalign{\smallskip}
B1   &$5 \times {10^7}$ &$1 \times {10^7}$&0.2 &0.006 &200  \\
B2   &$5 \times {10^7}$ &$1 \times {10^7}$&0.2 &0.01  &200  \\
B3   &$5 \times {10^7}$ &$1 \times {10^7}$&0.2 &0.02  &200  \\
B4   &$5 \times {10^7}$ &$1 \times {10^7}$&0.1 &0.01  &200  \\
B5   &$5 \times {10^7}$ &$1 \times {10^7}$&0.3 &0.01  &200  \\
B6   &$5 \times {10^7}$ &$5 \times {10^6}$&0.2 &0.01  &200  \\
B7   &$5 \times {10^7}$ &$2 \times {10^7}$&0.2 &0.01  &200  \\
B8   &$1 \times {10^8}$ &$1 \times {10^7}$&0.2 &0.006 &200  \\
B9  &$1.5 \times {10^8}$&$1 \times {10^7}$&0.2 &0.006 &200  \\
\hline\noalign{\smallskip}
\hline\noalign{\smallskip}
\end{tabular}
\end{center}

\begin{list}{}
\item\scriptsize{\textit{Note}. Column (2) is the Compton temperature of radiation; Columns (3) and (4) are the gas temperature and accretion rate at the outer boundary, respectively; Column (5) is LLAGNs luminosity; column (6) is the stellar velocity dispersion $\sigma$.}
\end{list}
\label{table1}
\end{table}

A-type models (models A1--A8) assume that the gas across the inner boundary of the computational domain freely falls to the circularization radius ($r_{\rm cir}$) and then is accreted like a hot accretion flow. In this case, the accretion rate is very low. Numerical simulations of hot accretion flow imply that the mass inflow rate decreases inwards insides $r_{\rm cir}$ due to the existence of outflows (Stone et al. 1999; Yuan, Wu \& Bu 2012). According to the simulation results, the net accretion rate ($\dot{M}_{\rm net}$) is given by $\dot{M}_{\rm net}=(10r_{\rm s}/r_{\rm cir})^{0.5}\dot{M}$ (Yuan, Bu \& Wu 2012; Yuan et al. 2015). After getting the $\dot{M}_{\rm net}$ from $\dot{M}$, we are then able to calculate the luminosity of LLAGNs by ${L_{\rm{X}}} = \epsilon {{\dot M}_{{\rm{net}}}}{c^2}$, where the radiative efficiency $\epsilon$ is given by
\begin{equation}
\epsilon  = \left\{ {\begin{array}{*{20}{c}}
{0.1,(5.3 \times {{10}^{ - 3}} < {{\dot M}_{{\rm{net}}}}/{{\dot M}_{{\rm{Edd}}}})}\\
{0.17{{(\frac{{100{{\dot M}_{{\rm{net}}}}}}{{{{\dot M}_{{\rm{Edd}}}}}})}^{1.12}},(3.3 \times {{10}^{ - 3}} < {{\dot M}_{{\rm{net}}}}/{{\dot M}_{{\rm{Edd}}}} < 5.3 \times {{10}^{ - 3}})}\\
{0.055{{(\frac{{100{{\dot M}_{{\rm{net}}}}}}{{{{\dot M}_{{\rm{Edd}}}}}})}^{0.076}},(2.9 \times {{10}^{ - 5}} < {{\dot M}_{{\rm{net}}}}/{{\dot M}_{{\rm{Edd}}}} < 3.3 \times {{10}^{ - 3}})}\\
{1.58{{(\frac{{100{{\dot M}_{{\rm{net}}}}}}{{{{\dot M}_{{\rm{Edd}}}}}})}^{0.65}},({{\dot M}_{{\rm{net}}}}/{{\dot M}_{{\rm{Edd}}}} \le 2.9 \times {{10}^{ - 5}})}
\end{array}} \right.
\end{equation}
(Xie \& Yuan 2012). Therefore, $L_{\rm X}$ and $\dot{M}$ are coupled in the A-type models. When the $\dot{M}$ in our models is given, the LLAGN luminosity is calculated and then the Eddington ratio ($f_{\rm X}$) is determined. In addition, if one chooses a different $r_{\rm cir}$, the calculated LLAGN luminosity would be changed. The effect of a different luminosity on the solution will be discussed in section 3.2.

B-type models (models B1--B9) assume that although all of the gases are still able to flow through the inner boundary, some of them will not reach the circularization radius due to some mechanism, such as the wind feedback from LLAGNs. Numerical simulations found that strong winds exist in hot accretion flows (Yuan, Bu \& Wu 2012; Yuan et al. 2015) and the winds could play an important role in the mechanical feedback of LLAGNs (e.g. Mou et al. 2014; Bu \& Yang 2019). Yuan et al. (2015) found that the wind mass flux is distributed within $\theta\sim30^{\rm o}$--$70^{\rm o}$ and $\theta\sim110^{\rm o}$--$150^{\rm o}$, where $\theta$ is the polar angle. The mechanical feedback of winds could prevent the gas from continuously falling. However, we do not consider the effect of winds on the gas within the computational domain. Then, under the previous assumption, the black hole still swallows gas at low accretion rate and keep low luminosity, but we could study the models with higher accretion rate at large radii. In order to avoid the area where the interacting of the winds and the falling gas occurs, a larger inner radius, i.e. 2000$r_{\rm s}$, is adopted in B-type models. Physical processes inside 2000$r_{\rm s}$ are neglected. Therefore, $L_{\rm X}$ and $\dot{M}$ are decoupled in the B-type models. $L_{\rm X}$ and $\dot{M}$ are taken as free parameters.

For each model, we have a set of model parameters, i.e. $T(r_{\rm out})$, ${T_{\rm{C}}}$, $\dot{M}$, and $L_{\rm X}$. As pointed out above, $L_{\rm X}$ and $\dot{M}$ are coupled in A-type models while they are decoupled in B-type models. In the A-type models, $\dot{M}$ is a free parameter while both $L_{\rm X}$ and $\dot{M}$ in the B-type models are free parameters. Since our models aim at LLAGNs, whose luminosity should not exceed 2\% $L_{\rm Edd}$, the X-ray luminosity of models should be less then 0.02 $L_{\rm Edd}$. Observations imply that the Compton temperature of LLAGNs radiation is in the range of 5--15$\times10^{7}$ K. Therefore, we set ${T_{\rm{C}}}$ to be $5 \times {10^7}$ K, $1 \times {10^8}$ K and $1.5 \times {10^8}$ K, respectively.

\subsection{Methods}
For each model, when $T(r_{\rm out})$, ${T_{\rm{C}}}$, $\dot{M}$, and $L_{\rm X}$ are given, we can then calculate a steady state solution by solving equations (1), (2), and (4). We can reduce these equations to be two first-order differential equations (see appendix for detailed derivations), which are given by
\begin{equation}
\begin{aligned}
\frac{{{\rm{d}}\ln \rho }}{{{\rm{d}}\ln r}} = & - \frac{{2{\mathcal{M}^2}}}{{{\mathcal{M}^2} - 1}}\\
 &+ \frac{\rho }{{\gamma ({\mathcal{M}^2} - 1)p}}[\frac{{G{M_{{\rm{BH}}}}}}{r}(\frac{{{r^2}}}{{{{(r - {r_{\rm{s}}})}^2}}} - {f_{\rm{X}}}) + {\sigma ^2}]\\
&+ \frac{{4\pi {r^3}\rho (\gamma  - 1)}}{{({\mathcal{M}^2} - 1)p\dot M\gamma }}{n^2}S
\end{aligned}
\end{equation}
and
\begin{equation}
\begin{aligned}
\frac{{{\rm{d}}\ln T}}{{{\rm{d}}\ln r}} = & - \frac{{2{\mathcal{M}^2}(\gamma  - 1)}}{{{\mathcal{M}^2} - 1}}\\
 &+ \frac{{\rho (\gamma  - 1)}}{{\gamma ({\mathcal{M}^2} - 1)p}}[\frac{{G{M_{{\rm{BH}}}}}}{r}(\frac{{{r^2}}}{{{{(r - {r_{\rm{s}}})}^2}}} - {f_{\rm{X}}}) + {\sigma ^2}]\\
 &+ \frac{{4\pi {r^3}\rho (\gamma  - 1)(\gamma {\mathcal{M}^2} - 1)}}{{({\mathcal{M}^2} - 1)p\dot{M}\gamma }}{n^2}S,
\end{aligned}
\end{equation}
where $\mathcal{M}$ ($=v^2\rho/\gamma p$) is the \textit{Mach} number of the accreting gas. We use the shooting method to solve equations (11)--(12), where a transonic solution is expected to be physical. At the outer boundary, when $T(r_{\rm out})$ and $\rho(r_{\rm out})$ are given, the Cash-Karp Runge-Kutta method with adaptive stepsize control (Press et al. 1992) is used to integrate from the outer boundary to the inner boundary. Here, we take $\dot{M}$ and $T(r_{\rm out})$ as a set of parameters. When $T(r_{\rm out})$ and $\dot{M}$ are given, we can change the density ($\rho(r_{\rm out})$) by adjusting the \textit{Mach} number at the outer boundary. However, due to the singularity at sonic point, there is a critical Mach number ($\mathcal{M}_{\rm c}$) at the outer boundary for a sets of $T(r_{\rm out})$ and $\dot{M}$ (Krolik \& London 1983; Mathews \& Guo 2012). When the \textit{Mach} number at the outer boundary is set to be less than $\mathcal{M}_{\rm c}$, the solution is not transonic. When the \textit{Mach} number at the outer boundary is set to be larger than $\mathcal{M}_{\rm c}$, the singularity at the sonic point prevents us from integrating inwards. When the \textit{Mach} number at the outer boundary is set to equal $\mathcal{M}_{\rm c}$, the transonic solution is obtained by integrating inwards. In general, for a given $\dot{M}$ and $T(r_{\rm out})$, we need to search the eigenvalue of $\mathcal{M}_{\rm c}$ at the outer boundary. When the eigenvalue of $\mathcal{M}_{\rm c}$ is got, we can calculate the inward velocity at the outer boundary based on $\mathcal{M}_{\rm c}$ and $T(r_{\rm out})$, and then obtain the gas density $\rho(r_{\rm out})$ at the outer boundary from $\dot{M}$.

In both observations and the numerical simulations of galaxy formation and evolution, we often need to estimate the mass accretion rate from the gas density and temperature at the parsec scale. In A-type models, the X-ray luminosity ($L_{\rm X}$) and the accretion rate ($\dot{M}$) are coupled. When A-type models are used to predict the accretion rates, an iteration method applies to calculate the accretion rates. When the gas density and temperature at the outer boundary are given, we can assume an accretion rate and then calculate $\rho(r_{\rm out})$ using the above method. An accretion rate can be adjusted until the obtained $\rho(r_{\rm out})$ is equal to the given gas density. In B-type models, the X-ray luminosity and the accretion rate are decoupled. We can integrate Equations (11) and (12) from the gas density and temperature at the outer boundary to search the eigenvalue of $\mathcal{M}_{\rm c}$ at the outer boundary. Then, the gas velocity at the outer boundary can be calculated using $\mathcal{M}_{\rm c}$, $\rho(r_{\rm out})$, and $T(r_{\rm out})$, so that $\dot{M}$ can be got.

\section{RESULTS}
In this section, we will present the results of our calculations and discuss their physical properties. All of the models listed in table 1 are thermally stable.

For A-type models, we first examine the effect of different $\dot{M}$ and $T(r_{\rm out})$ with models A1--A5 and then we examine the effect of Compton temperature ($T_{\rm C}$) with models A6 and A7. Model A8 is a model without the bulge stellar potential. Similarly, we also examine different values of each parameter in B-type models.

\subsection{Analyse of A-type models}
In A-type models, we test three parameters, i.e. $T(r_{\rm out})$, $\dot{M}$, and $T_{\rm C}$. For models A1--A3, $T_{\rm C}$ and $T_{\rm out}$ are set to be $5\times10^7$ K and $10^7$ K, respectively, while $\dot{M}$ varies from 0.013 to 0.045 ${\dot{M}_{\rm Edd}}$, as given in Table 1. The solutions of A1--A3 are shown in Figure 1, where the Bondi solution is also given for comparison. In addition, Figure 2 gives their radiative heating/cooling rate. The Bondi solution depends on the gas temperature and density at infinity. We also use the density and temperature at the outer boundary in models A1--A3 to calculate the corresponding Bondi solutions of models A1--A3 For models A1--A3, since the change of accretion rate is not so significant, their corresponding Bondi solutions do not have an obvious difference. Besides, although the adiabatic index $\gamma$ is set to be 5/3, the Bondi solution is still able to become supersonic due to that Paczy\'{n}ski-Wiita potential is adopted, as shown in Figure 1. Figure 1 shows that the solutions of A1--A3 are significantly different from the Bondi solution. Compared with the Bondi solution, models A1--A3 have higher velocity and lower temperature as well as higher \textit{Mach} number because that they undergo significant Compton cooling at r $<$ 0.1pc. Especially, the solutions of A1--A3 become supersonic inside about 0.02 pc while the Bondi solution becomes transonic at smaller radii. Comparing models A1--A3, their dynamic difference is negligible while their thermal properties have obvious differences, especially inside 0.1 pc.

In order to understand the thermal properties of solutions, equation (4) is written as
\begin{equation}
\frac{{{\rm{d}}T}}{{{\rm{d}}r}} = {C_1}\frac{T}{\rho } \cdot \frac{{{\rm{d}}\rho }}{{{\rm{d}}r}} + {C_2}\frac{{{n^2}S}}{{\rho v}} ,
\end{equation}
where $C_{1}$ and $C_{2}$ are constant. In this equation, the first term on the right-hand side describes the effect of the adiabatic compressional heating while the second term describes the effect of the radiative heating/cooling. The first term implies that a higher gas temperature leads to a more important compressional heating. The second term divided by $dT/dr$ is roughly proportional to $\frac{n^2 S r}{e v}$, which is the ratio of the dynamical timescale (${\tau _{\rm acc}} = r/v$) to the radiative-heating/cooling timescale(${\tau _{\rm rad}} = e/{n^2}S$). When the falling gas moves faster inwards, the dynamical timescale becomes shorter and then the effect of radiative heating/cooling becomes relatively weak. When the falling gas becomes dense, the radiative timescale becomes shorter, and then the radiative heating/cooling becomes relatively more important. Figure 2 shows the ${\tau _{\rm acc}}/{\tau _{\rm rad}}$ values of four radiative heating/cooling processes. When ${\tau _{\rm acc}}/{\tau _{\rm rad}} > 1$ for some radiative heating/cooling process, the radiative heating/cooling process becomes important in models. This is helpful to understand thermal differences between different models.

Figures 1 and 2 also show that there are three stages of temperature change. (1) For the falling gas from the outer boundary to the location ($r_{1}$) of the first extreme value of temperature, the thermodynamic processes are dominated by bremsstrahlung cooling, which causes the gas temperature to decrease inwards. (2) After the falling gas goes across $r_{1}$, the adiabatic compressional heating becomes strong and overwhelms the bremsstrahlung cooling, which makes the gas temperature begin to increase. (3) When the temperature rises above the Compton temperature of radiation, the Compton scattering becomes a mechanism of radiative cooling and the Compton cooling rate gradually increases inwards. When the sum of radiative cooling balances the adiabatic compressional heating again, the gas temperature achieves the second extreme value. This value is a maximum value, whose location is given by $r_{2}$ in Figure 1. After the falling gas goes across $r_{2}$, the gas temperature decreases again.

As shown in Figure 1, models A1--A3 have approximately the same radial velocity and then their dynamical timescale also has the approximately same value. For the model with a higher accretion rate, its density ($\rho$) becomes higher at all radii, which makes the radiative heating/cooling relatively stronger. Therefore, the radiative cooling becomes more significant in model A3 than in models A1 and A2, and the temperature of model A3 begins to decreases at a larger radius and decreases faster than that of model A1.

\begin{figure*}
\scalebox{0.38}[0.38]{\rotatebox{0}{\includegraphics{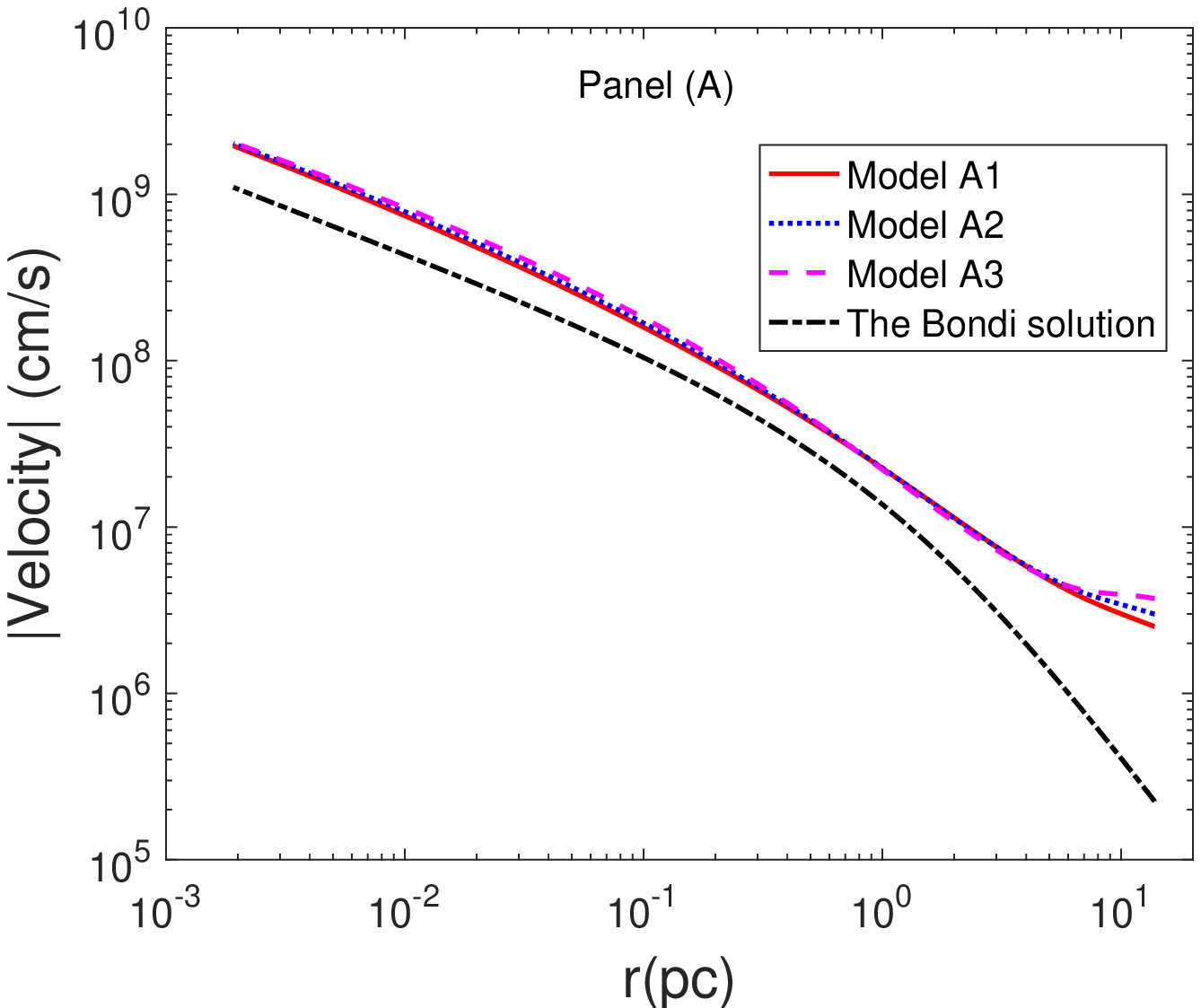}}}
\scalebox{0.38}[0.38]{\rotatebox{0}{\includegraphics{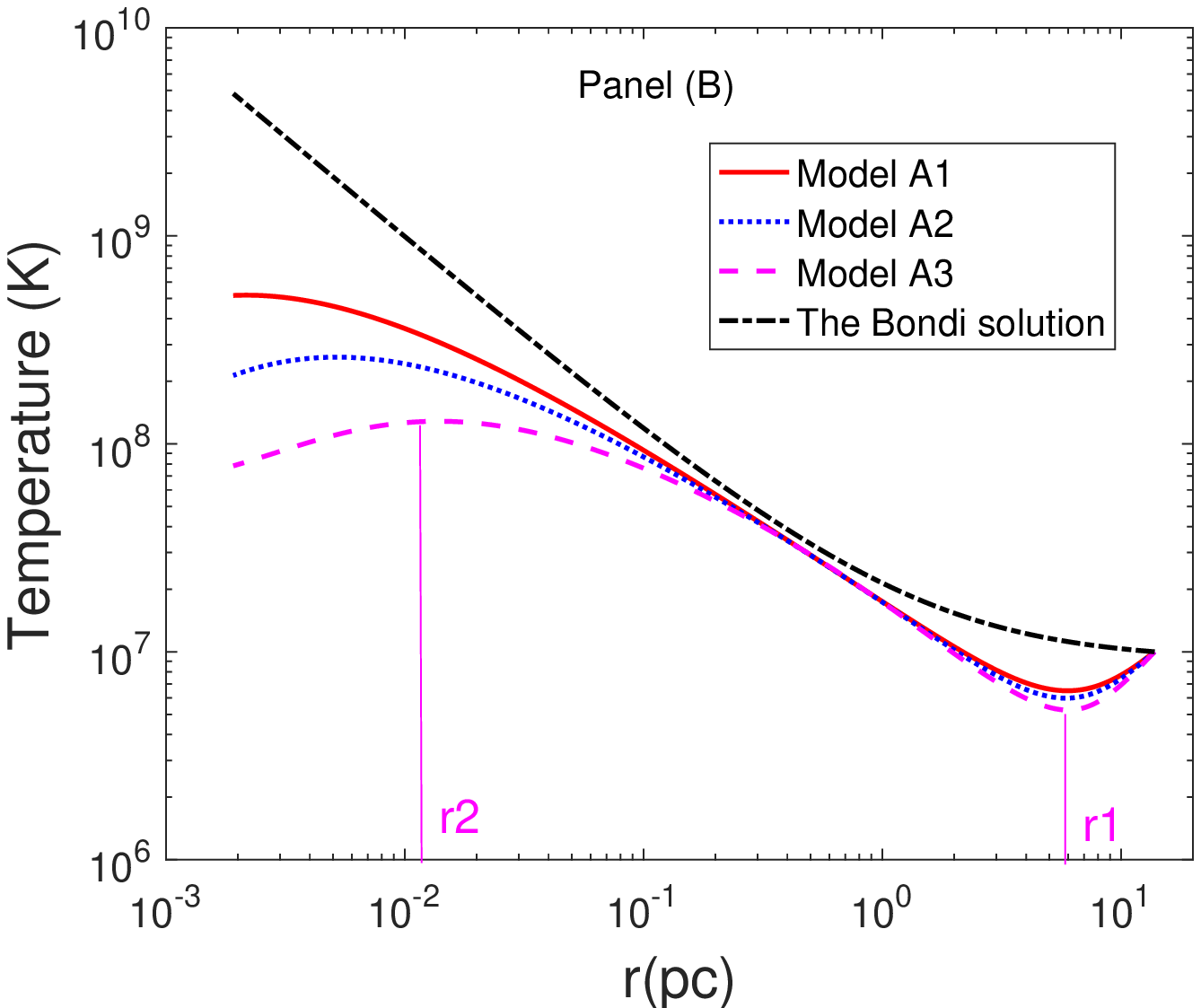}}}
\scalebox{0.38}[0.38]{\rotatebox{0}{\includegraphics{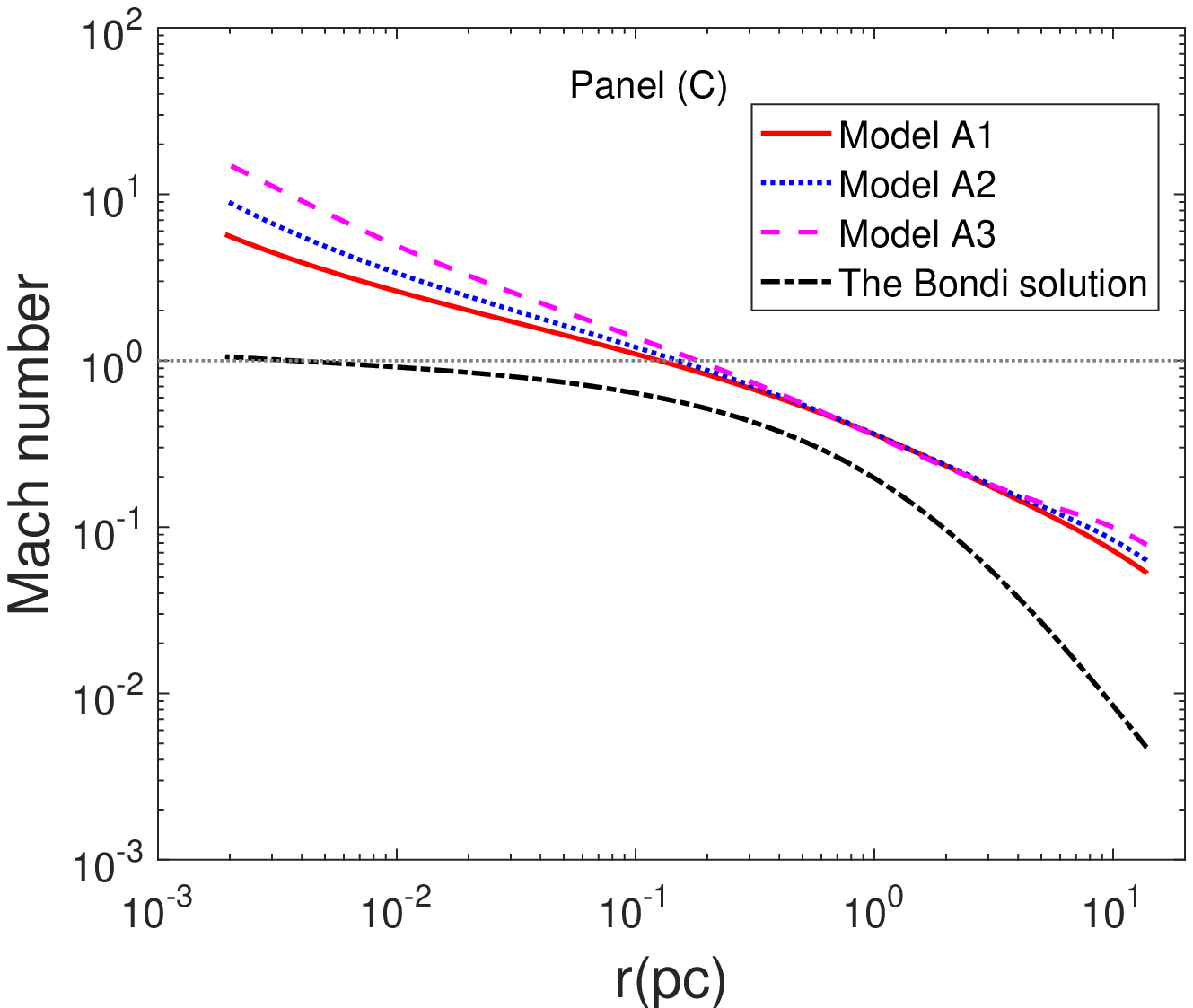}}}
\ \centering \caption{Radial dependent of velocity, temperature and \textit{Mach} number in models A1--A3.}\label{fig 1}
\end{figure*}

\begin{figure*}
\scalebox{0.38}[0.38]{\rotatebox{0}{\includegraphics{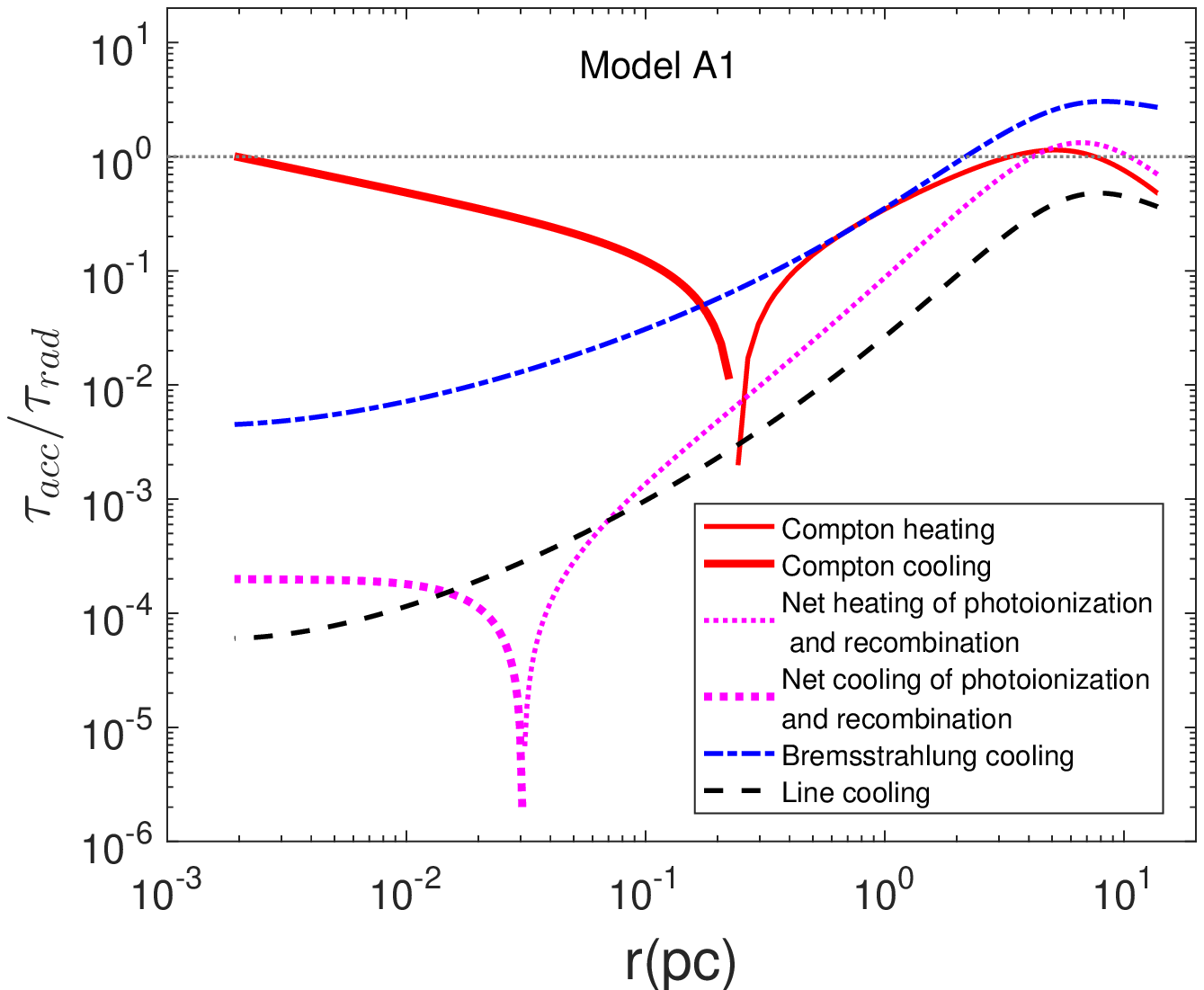}}}
\scalebox{0.38}[0.38]{\rotatebox{0}{\includegraphics{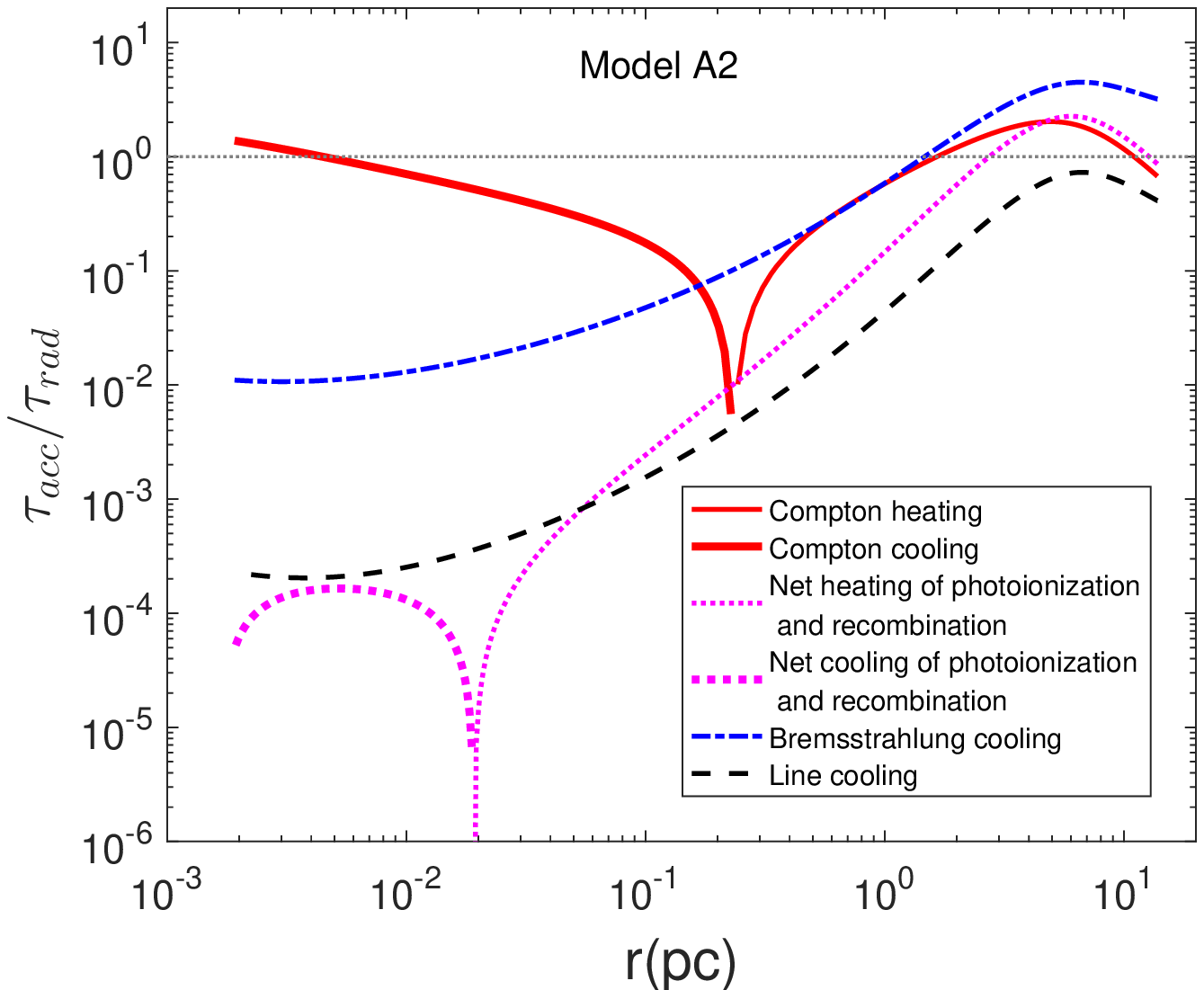}}}
\scalebox{0.38}[0.38]{\rotatebox{0}{\includegraphics{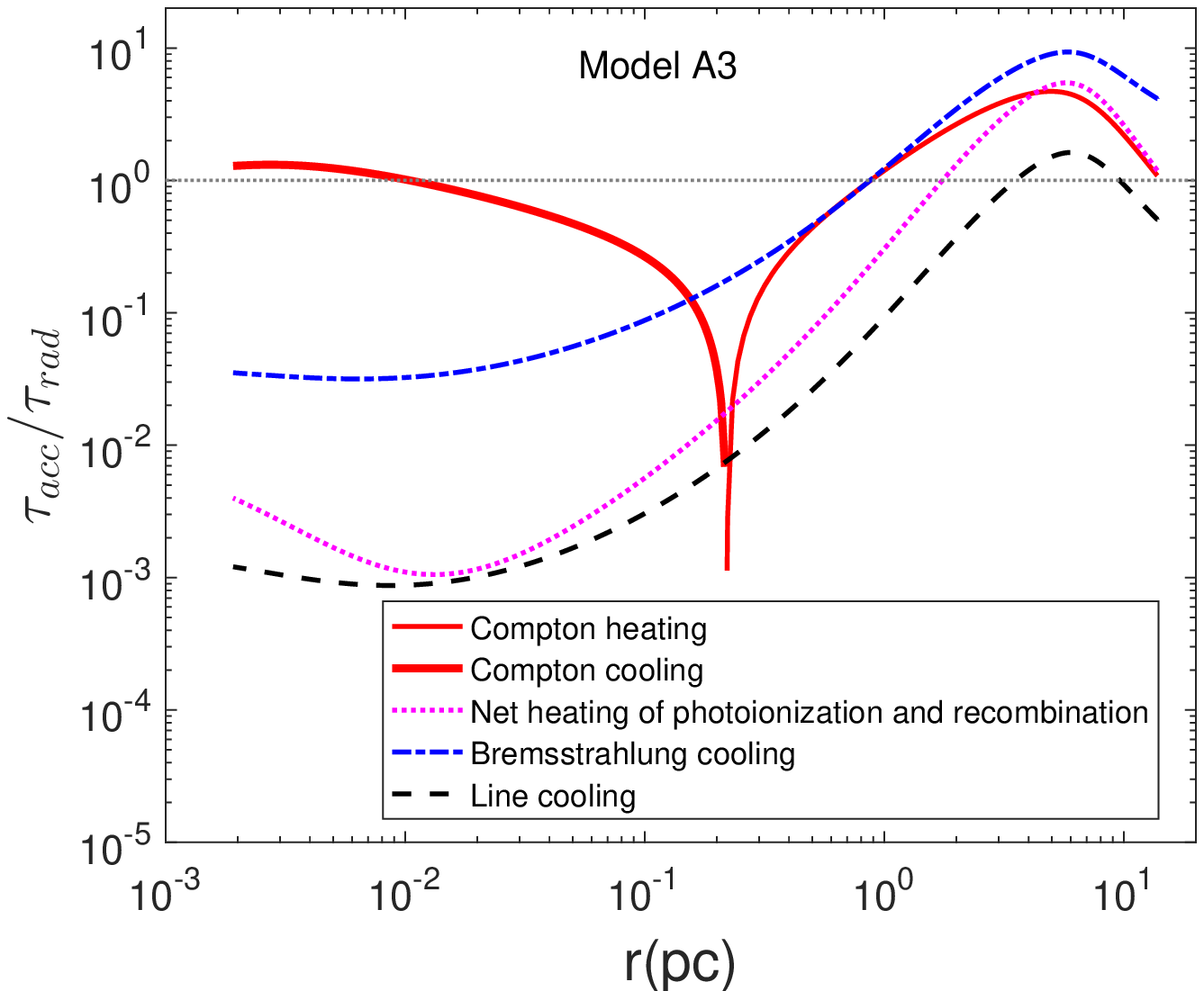}}}
\ \centering \caption{The ratio of the dynamical timescale (${\tau _{acc}}$) to the radiative heating/cooling timescale (${\tau _{rad}}$) of models A1, A2 and A3.}\label{fig 2}
\end{figure*}

\begin{figure*}
\scalebox{0.38}[0.38]{\rotatebox{0}{\includegraphics{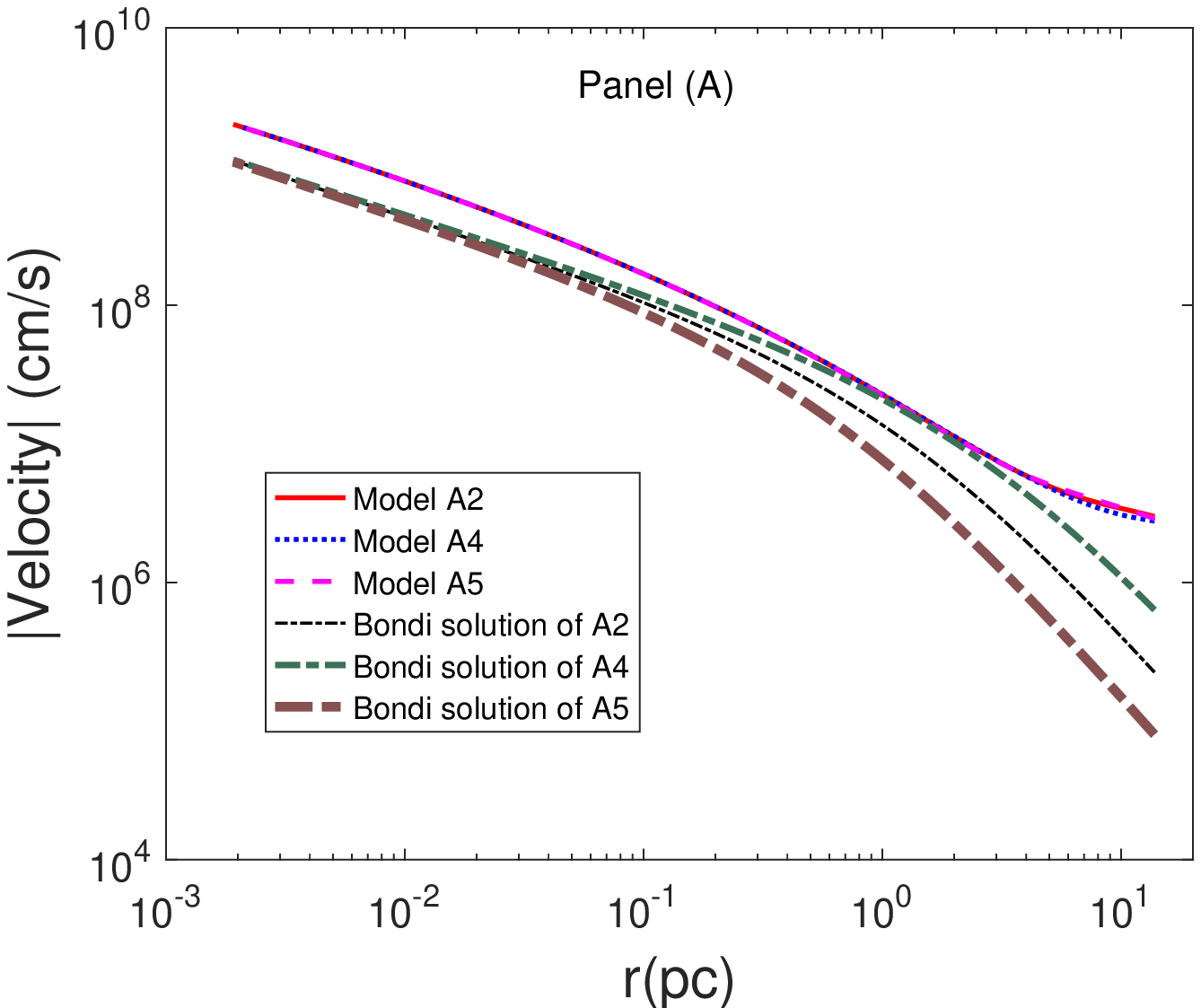}}}
\scalebox{0.38}[0.38]{\rotatebox{0}{\includegraphics{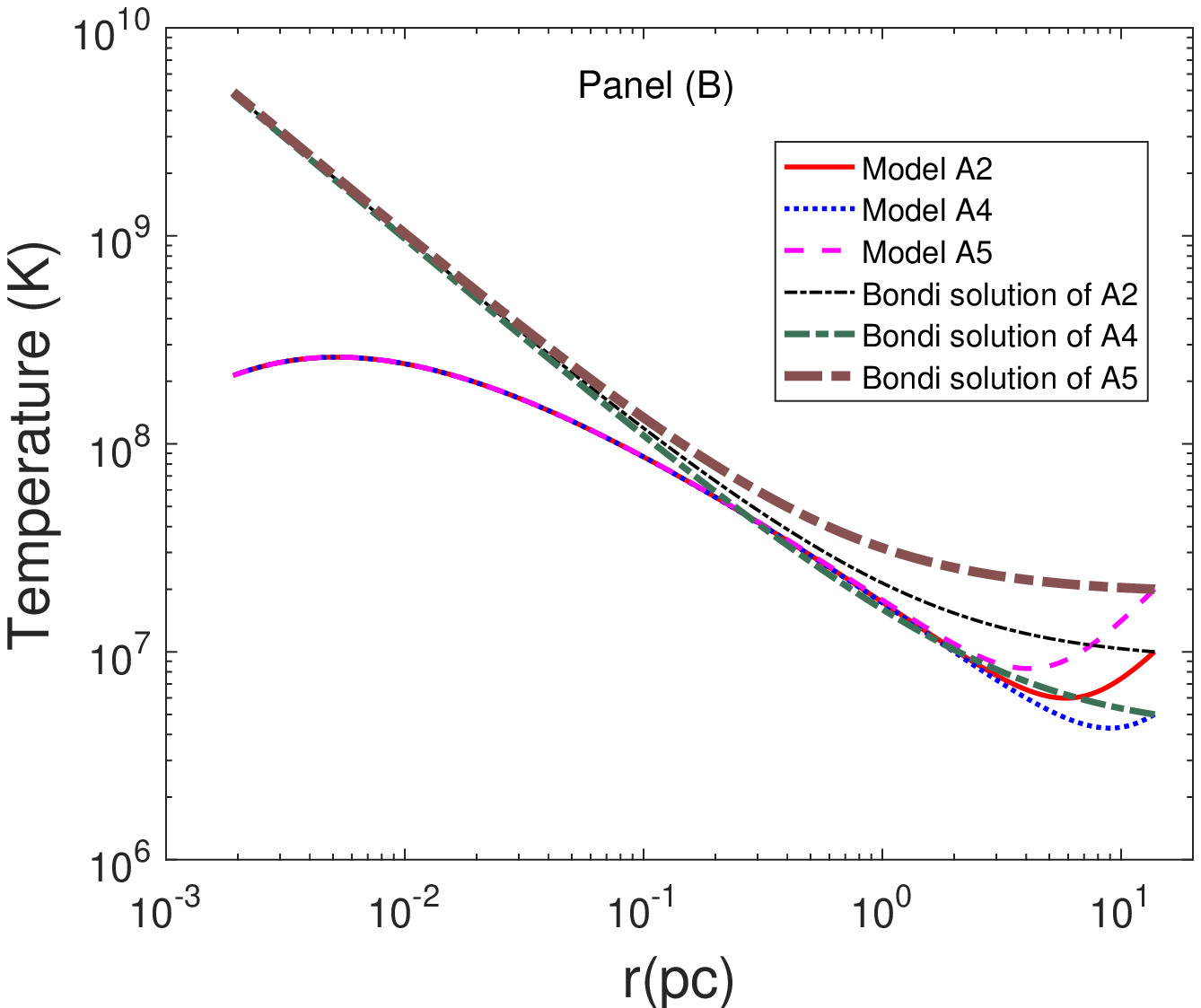}}}
\scalebox{0.38}[0.38]{\rotatebox{0}{\includegraphics{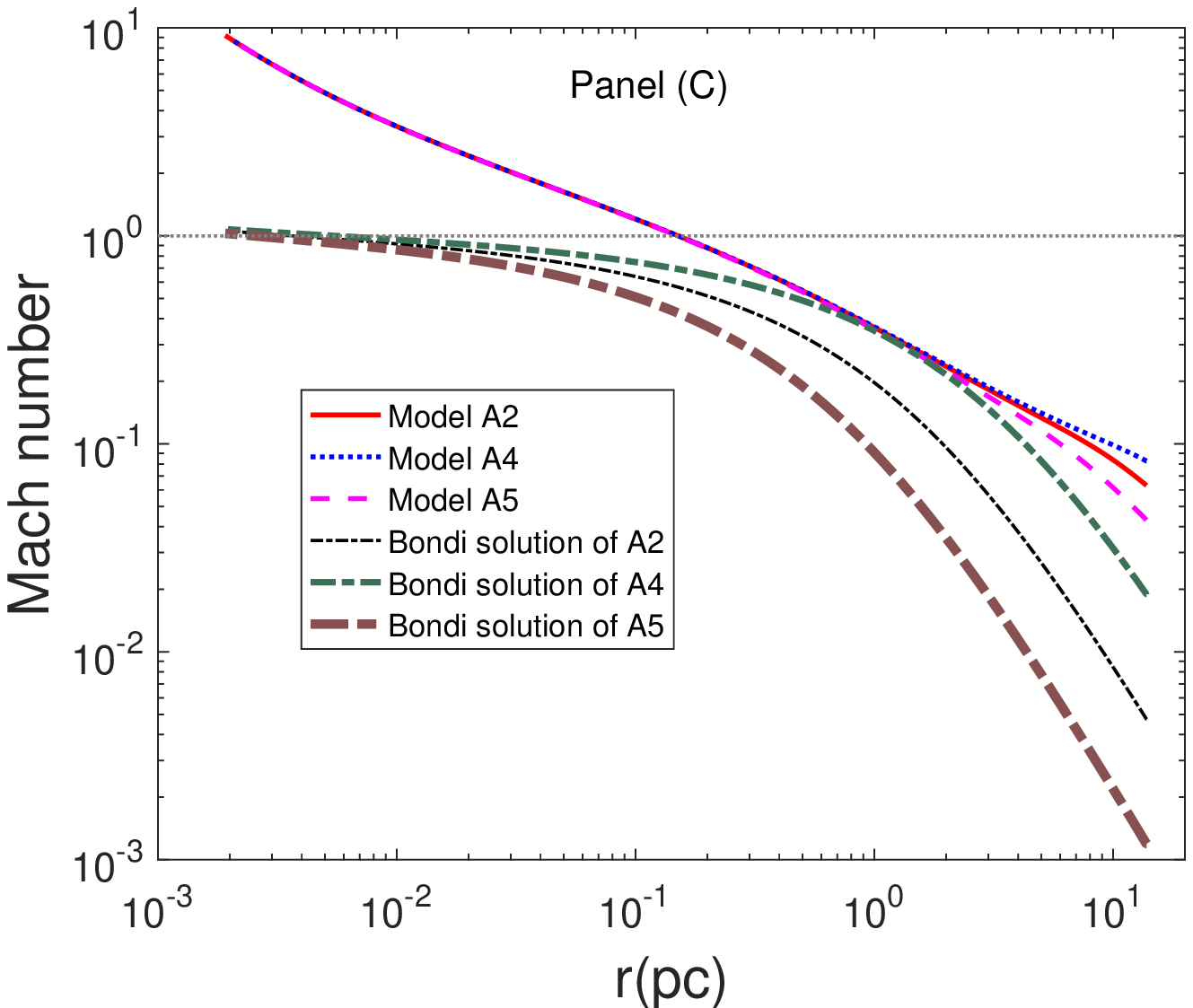}}}
\ \centering \caption{Radial dependent of velocity, temperature and \textit{Mach} number in models A2, A4, and A5.}\label{fig 3}
\end{figure*}

Models A2, A4, and A5 are used to test the effect of $T(r_{\rm out})$ and Figure 3 gives the solutions of these models. Due to the difference of gas temperature at the outer boundary, Bondi solutions of these three models have obvious differences. Figure 3 also shows corresponding Bondi solutions for each model. As shown in Figure 3, the three models have the almost same radial velocity. Because their accretion rate also has the same value, their density is almost the same at all radii. In this case, a higher temperature at the outer boundary strengthens the bremsstrahlung cooling. As a result, the gas temperature of model A5 decreases faster at large radii, compared with models A2 and A4. When the gas goes across a location of the minimum temperature and continuously falls inwards, compressional heating makes the increment of internal energy much larger than initial internal energy. This makes the difference in gas temperature gradually becomes small at small radii.

\begin{figure*}
\scalebox{0.38}[0.38]{\rotatebox{0}{\includegraphics{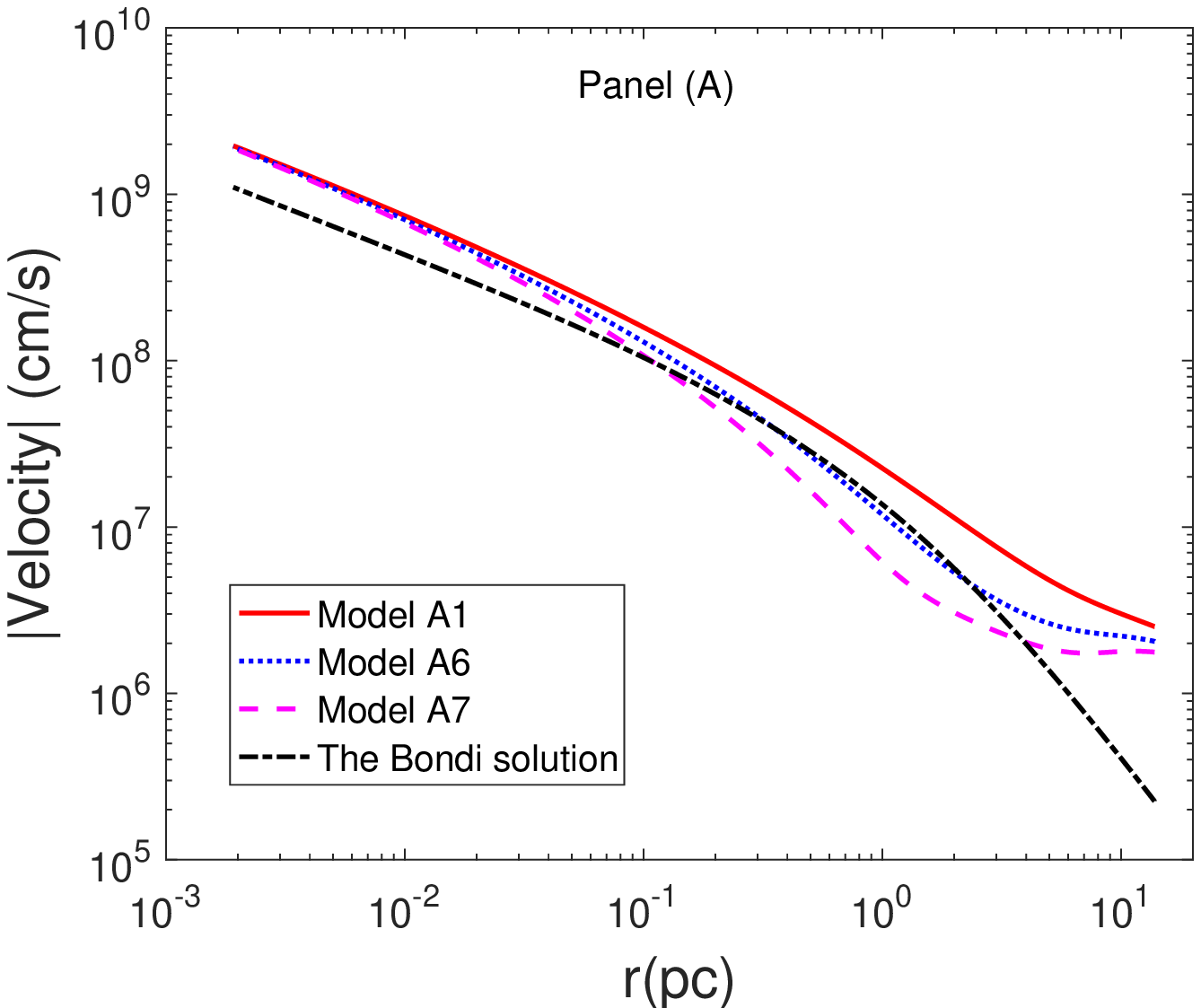}}}
\scalebox{0.38}[0.38]{\rotatebox{0}{\includegraphics{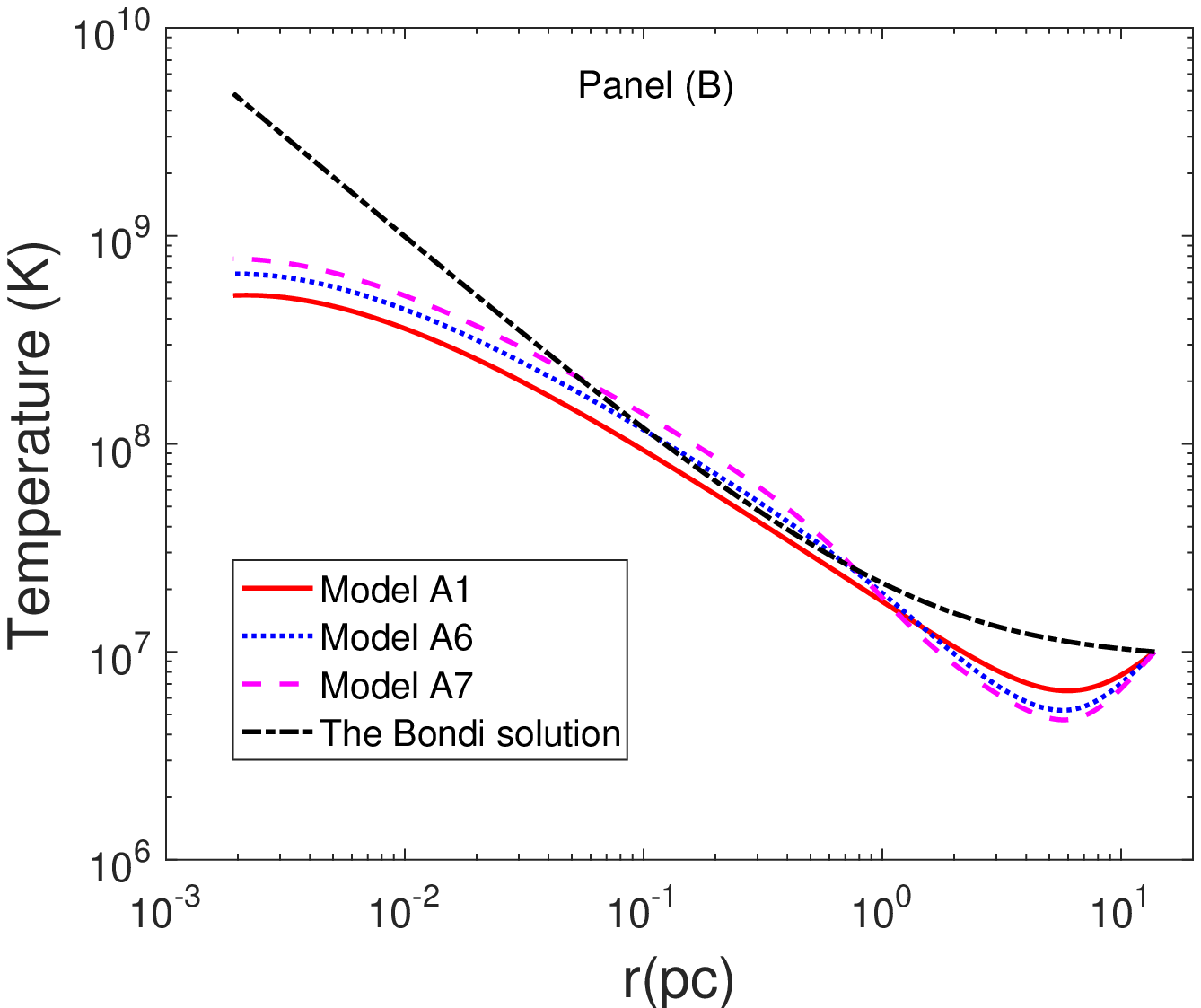}}}
\scalebox{0.38}[0.38]{\rotatebox{0}{\includegraphics{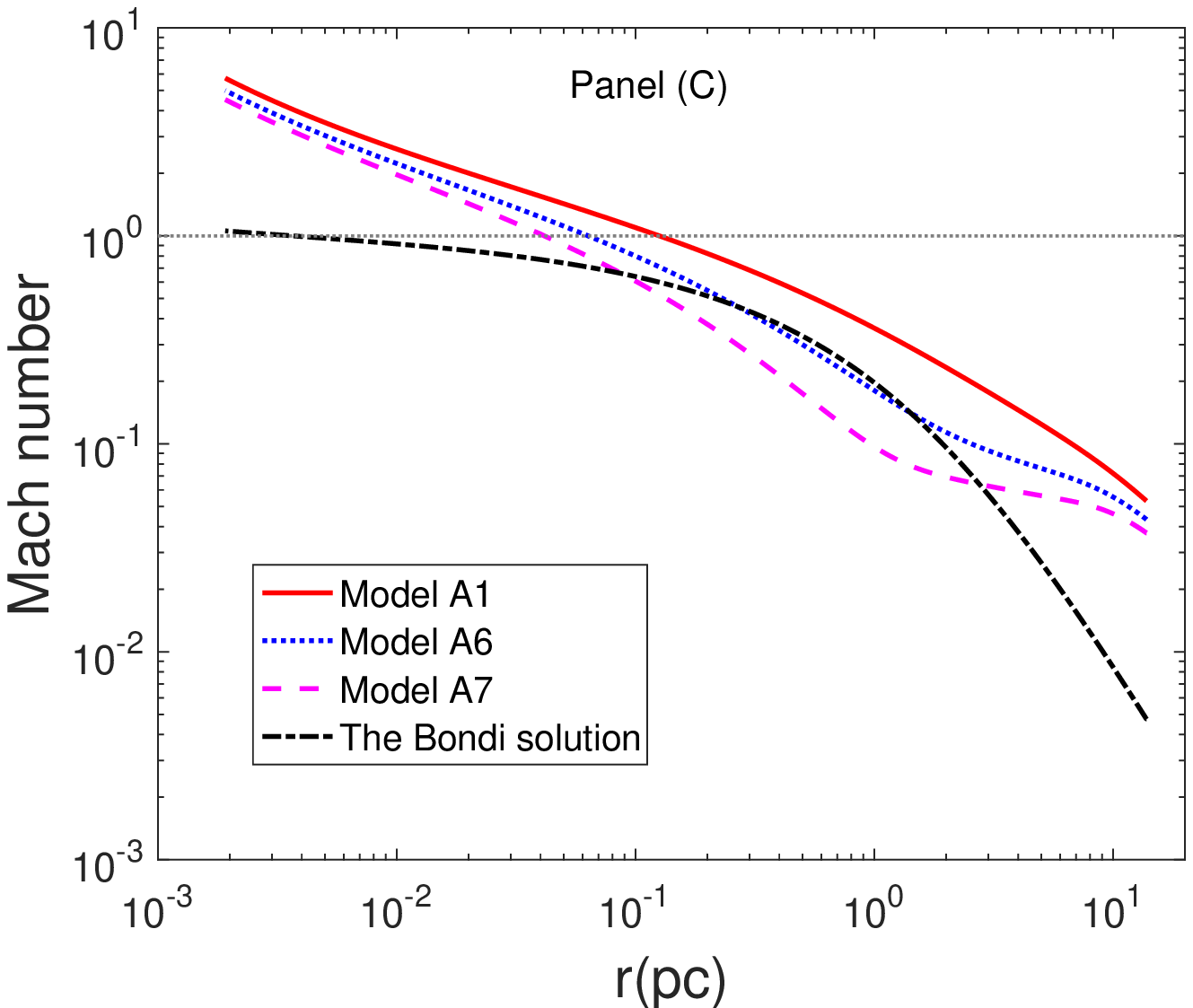}}}
\ \centering \caption{Radial dependent of velocity, temperature and \textit{Mach} number in models A1, A6, and A7.}\label{fig 4}
\end{figure*}

Models A1, A6, and A7 are used to test the effect of $T_{\rm C}$ and Figure 4 gives their solutions. As shown in Figure 4, the solutions are different in both dynamics and thermodynamics. In the dynamical properties, at the large radii, the gas falls slower in the model with higher Compton temperature. At the small radii, the three models almost have a close radial velocity. In the thermal properties, at the radii of $>$1 pc, the ${\tau _{\rm acc}}/{\tau _{\rm rad}}$ is larger in the model with lower velocity and then bremsstrahlung cooling has a longer time to make the gas slightly cooler. In the region where gas temperature is higher than $T_{\rm C}$, higher $T_{\rm C}$ means weaker Compton radiation cooling. Therefore, the maximum value of gas temperature in the model with higher $T_{\rm C}$ is higher than that in the model with lower $T_{\rm C}$.

\subsection{Analyse of B-type Models}

\begin{figure*}
\scalebox{0.38}[0.38]{\rotatebox{0}{\includegraphics{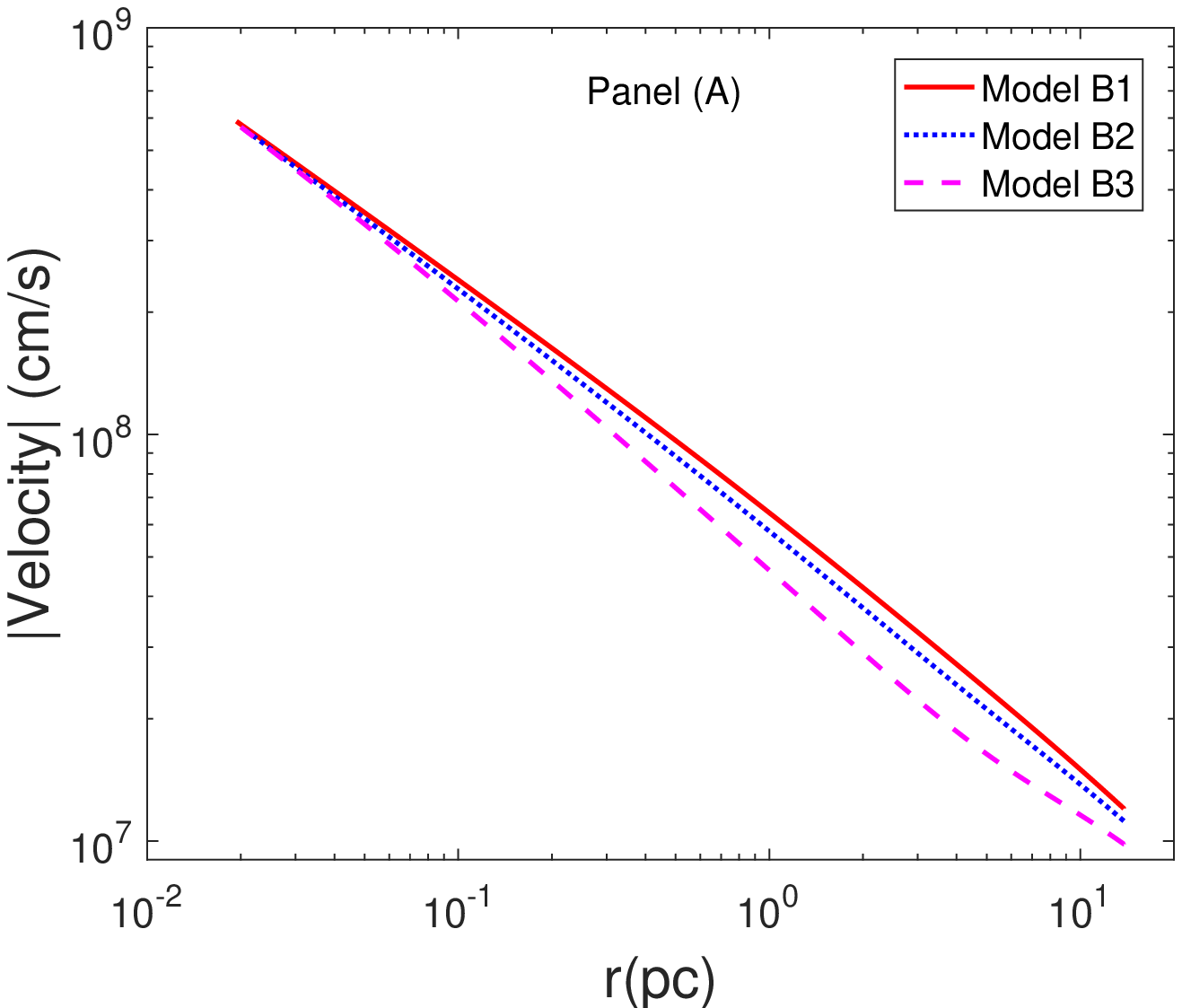}}}
\scalebox{0.38}[0.38]{\rotatebox{0}{\includegraphics{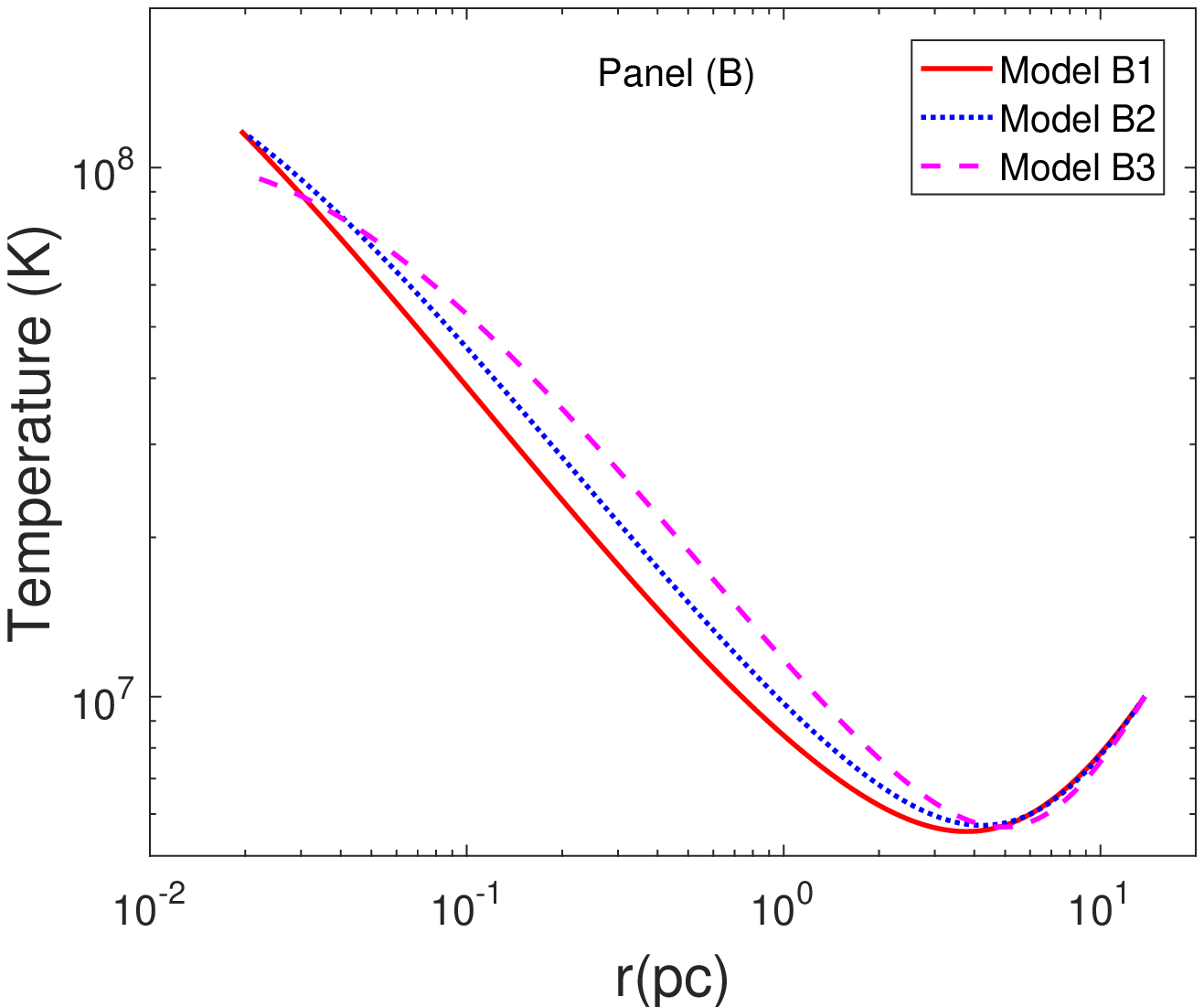}}}
\scalebox{0.38}[0.38]{\rotatebox{0}{\includegraphics{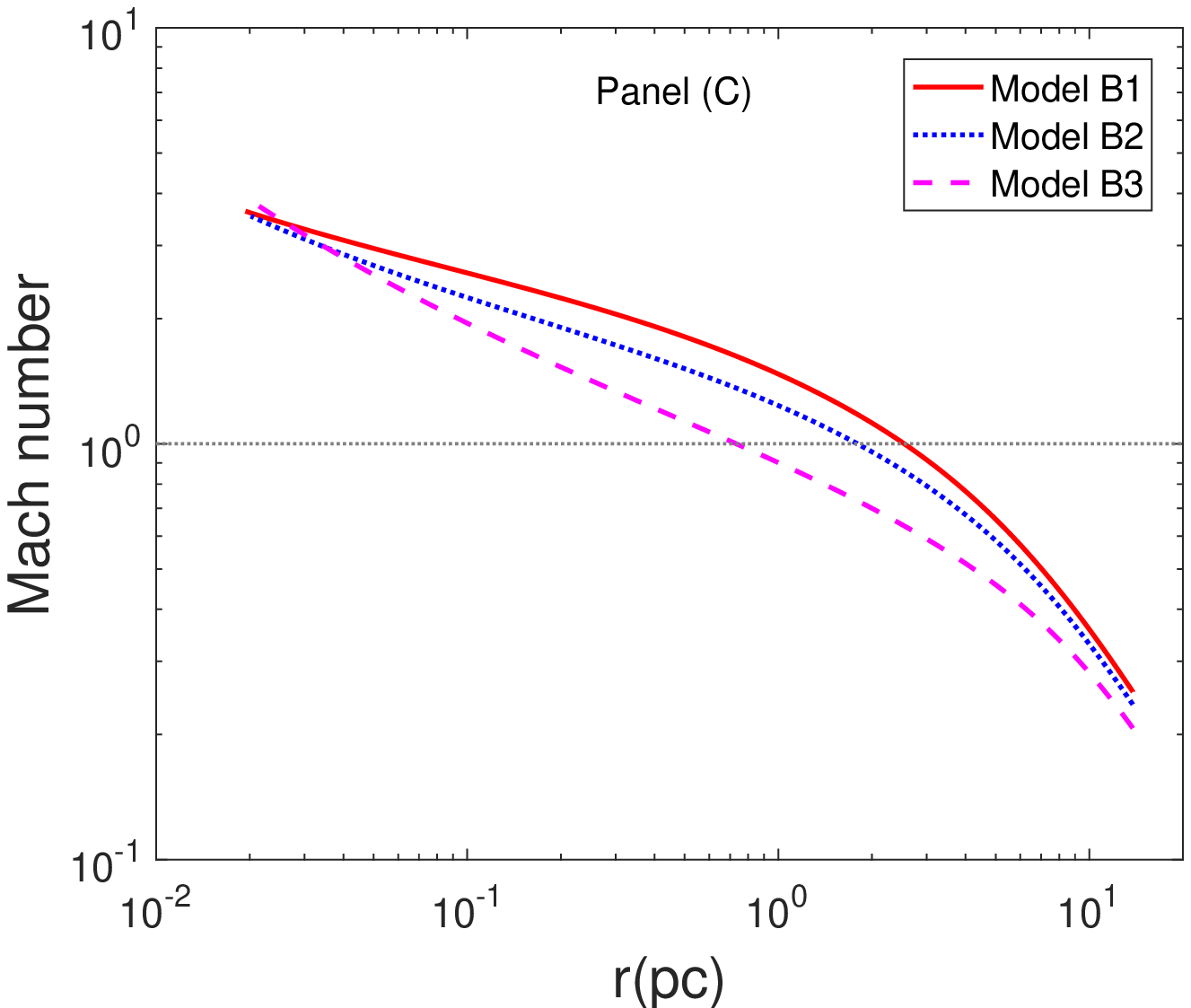}}}
\ \centering \caption{Radial dependent of velocity, temperature and \textit{Mach} number in models B1, B2, and B3. }\label{fig 5}
\end{figure*}

\begin{figure*}
\scalebox{0.38}[0.38]{\rotatebox{0}{\includegraphics{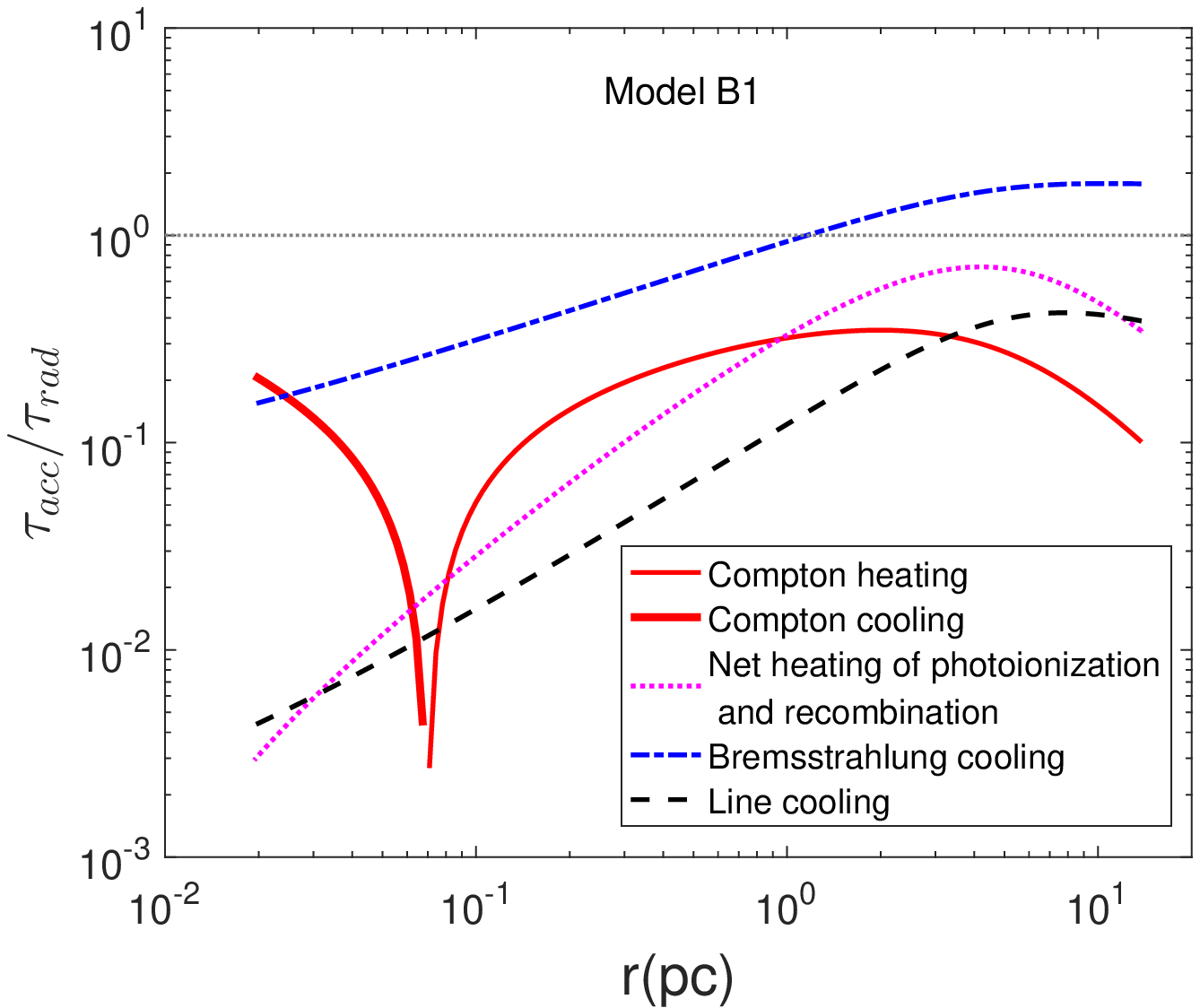}}}
\scalebox{0.38}[0.38]{\rotatebox{0}{\includegraphics{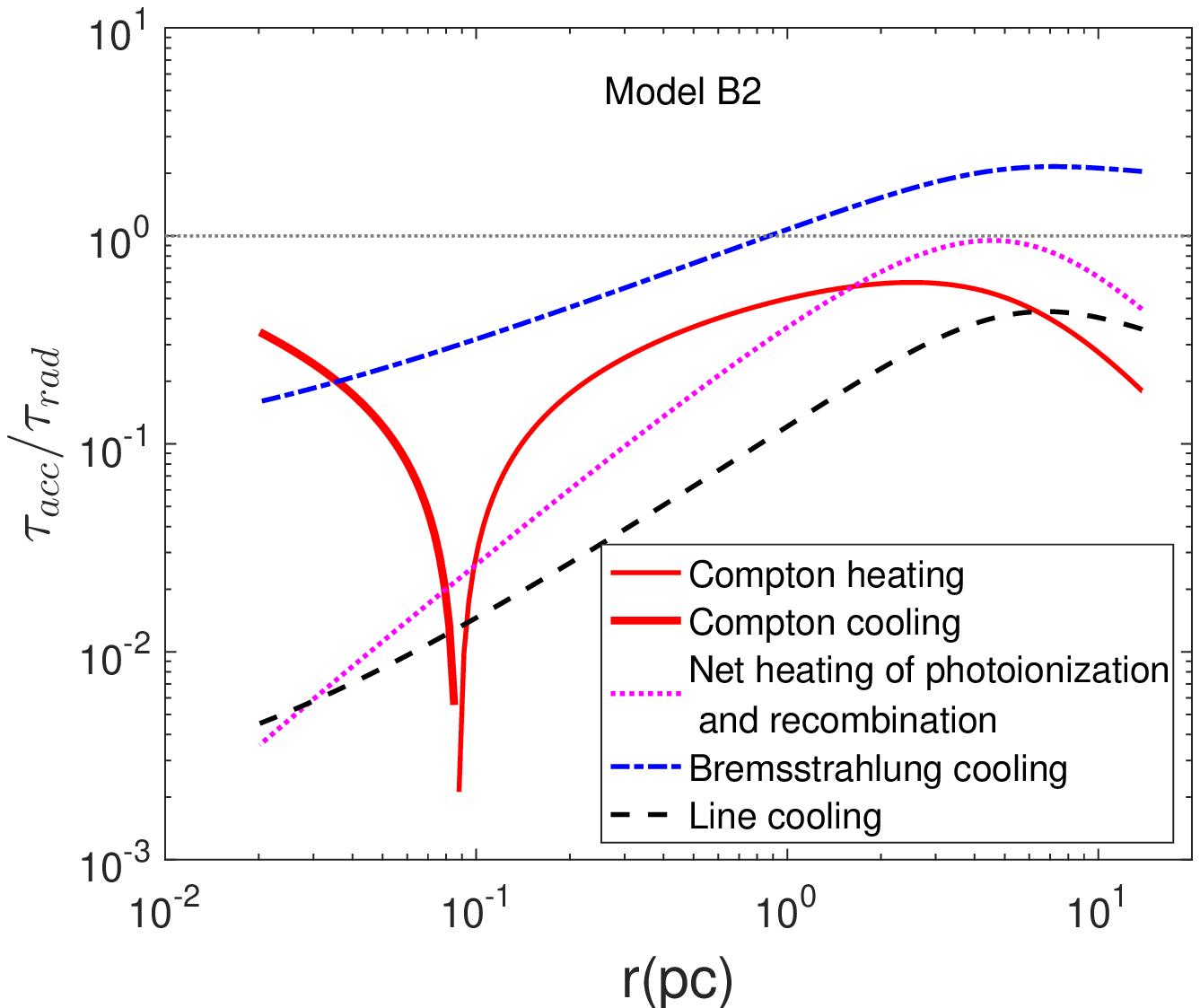}}}
\scalebox{0.38}[0.38]{\rotatebox{0}{\includegraphics{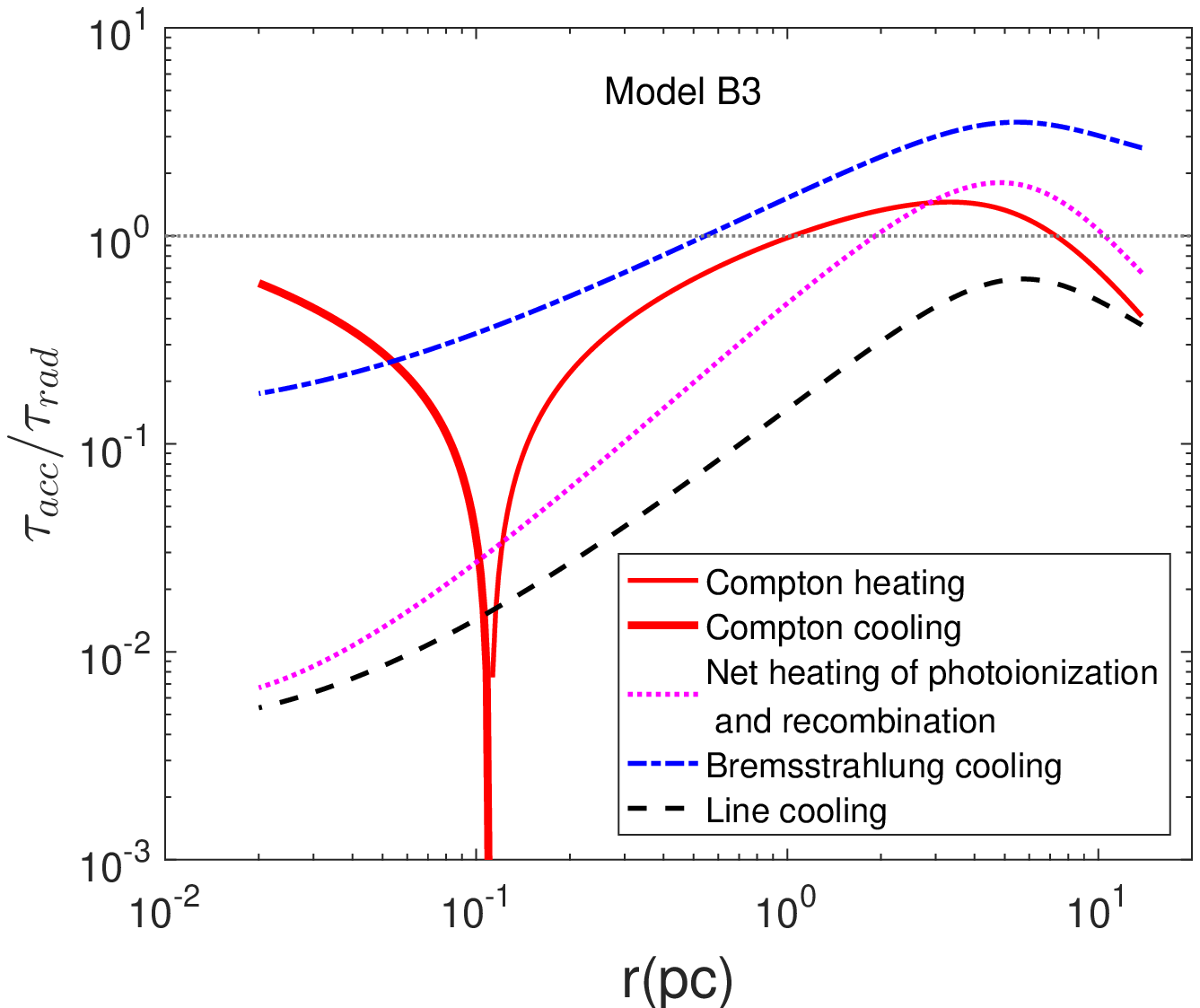}}}
\ \centering \caption{The ratio of the dynamical timescale (${\tau _{acc}}$) to the radiative heating/cooling timescale (${\tau _{rad}}$) of models B1, B2 and B3.}\label{fig 6}
\end{figure*}

In B-type models, luminosity is treated as a free parameter, which requires that the given $\dot{M}$ in Table 1 is high enough to produce at least the given $L_{\rm X}$ in Table 1. Then, we can separately test the effects of different accretion rates and luminosity. For example, models B1--B3 with the same accretion rate are used to test the effects of different luminosity. Figure 5 gives their solutions and Figure 6 shows the radial dependence of ${\tau _{\rm acc}}/{\tau _{\rm rad}}$ for these three models. Panel (A) in Figure 5 shows that, for the model B3 where a higher luminosity is adopted, the falling gas moves inwards slower by a factor of $\sim$20$\%$ than that of model B1 at large radii ($\sim$10pc), which means that the accretion time becomes longer by the same factor and then the falling gas has a longer time to cool. As a result, at large radii where the Bremsstrahlung cooling is dominant, the gas temperature of model B3 decreases faster by a factor of $\sim$20$\%$ than the gas temperature of model B1. At medium radii ($\sim$0.1--2pc), compared with model B1, the radiative heating of model B3 is relatively stronger by a factor of $\sim$40$\%$, which makes the gas temperature increase faster. At small radii ($\sim <$0.1pc), Compton heating is turned into Compton cooling and the radiative heating/cooling is dominated by Compton cooling. Compared with model B1, the Compton cooling of model B3 becomes stronger by a factor of about $\sim$2.2, which makes the increase of gas temperature becomes slower so that in model B3, the gas temperature at the inner boundary is lower than the other models.

We compare three models with different accretion rate, i.e. models B2, B4 and B5, and give results in Figure 7. Panels (A) and (B) in Figure 7 show that, compared to model B4 whose accretion rate is smaller than that of model B5, both density and radial velocity in model B5 have a higher value at the outer boundary. Compared to model B4, in model B5, the inwards increasing of radial velocity becomes slower and the inwards increasing of density becomes faster. In this case, in model B5, the inwards decreasing of accretion timescale becomes slow while the inward decreasing of radiative-cooling timescale becomes fast. With the decrease of radius, the ratio of the $\tau_{\rm acc}/\tau_{\rm rad}$ value of model B5 to the $\tau_{\rm acc}/\tau_{\rm rad}$ value of model B4 increases. Therefore, with the falling of gas, radiative cooling in the model with high accretion rate becomes relatively significant, compared to the model with low accretion rate, which causes that the inward increasing of gas temperature becomes slower in the model with higher accretion rate. As shown in panel (C), the gas temperature of model B5 is relatively lower at $r<\sim10$pc.

\begin{figure*}
\scalebox{0.5}[0.5]{\rotatebox{0}{\includegraphics{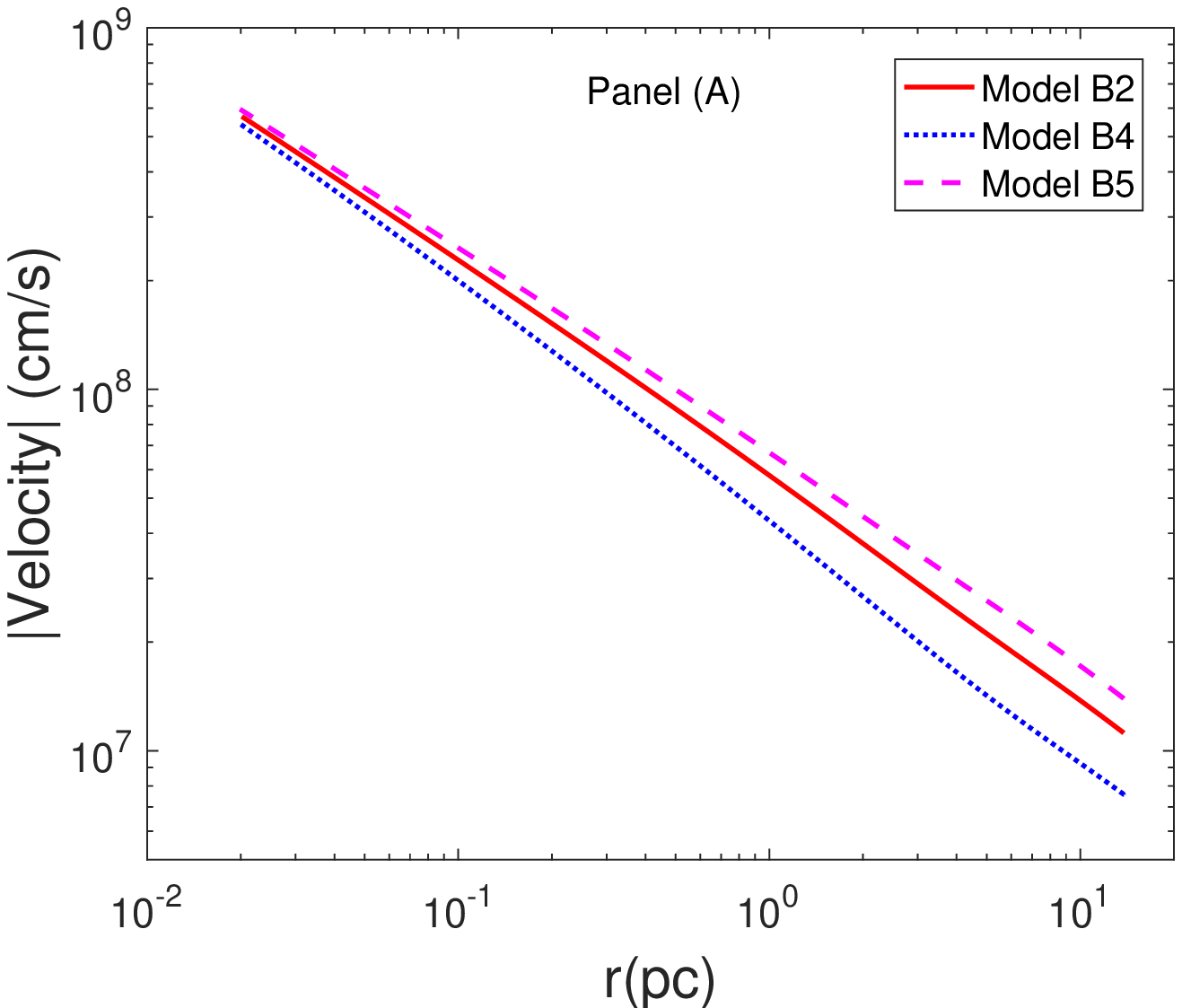}}}
\scalebox{0.5}[0.5]{\rotatebox{0}{\includegraphics{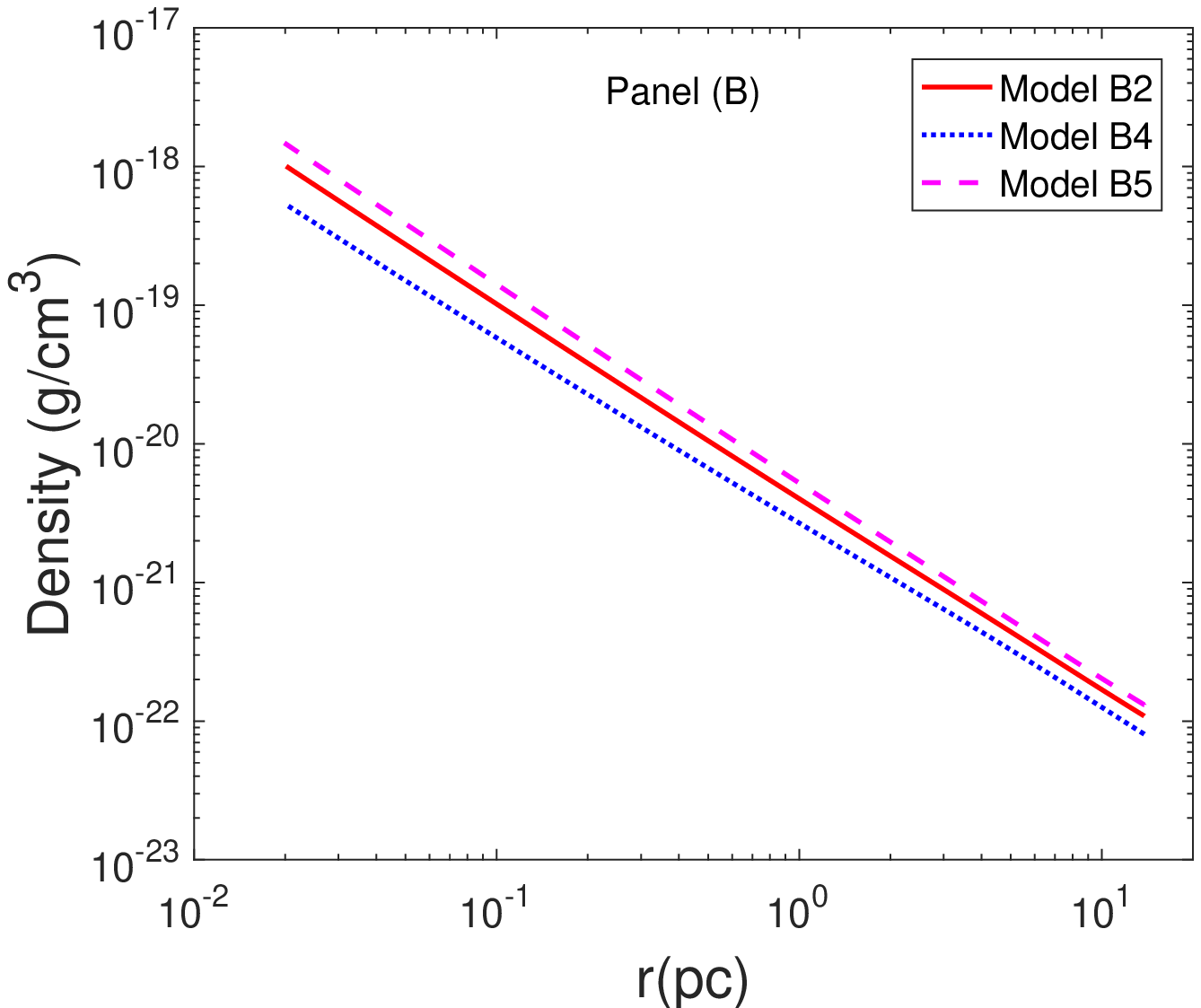}}}
\scalebox{0.5}[0.5]{\rotatebox{0}{\includegraphics{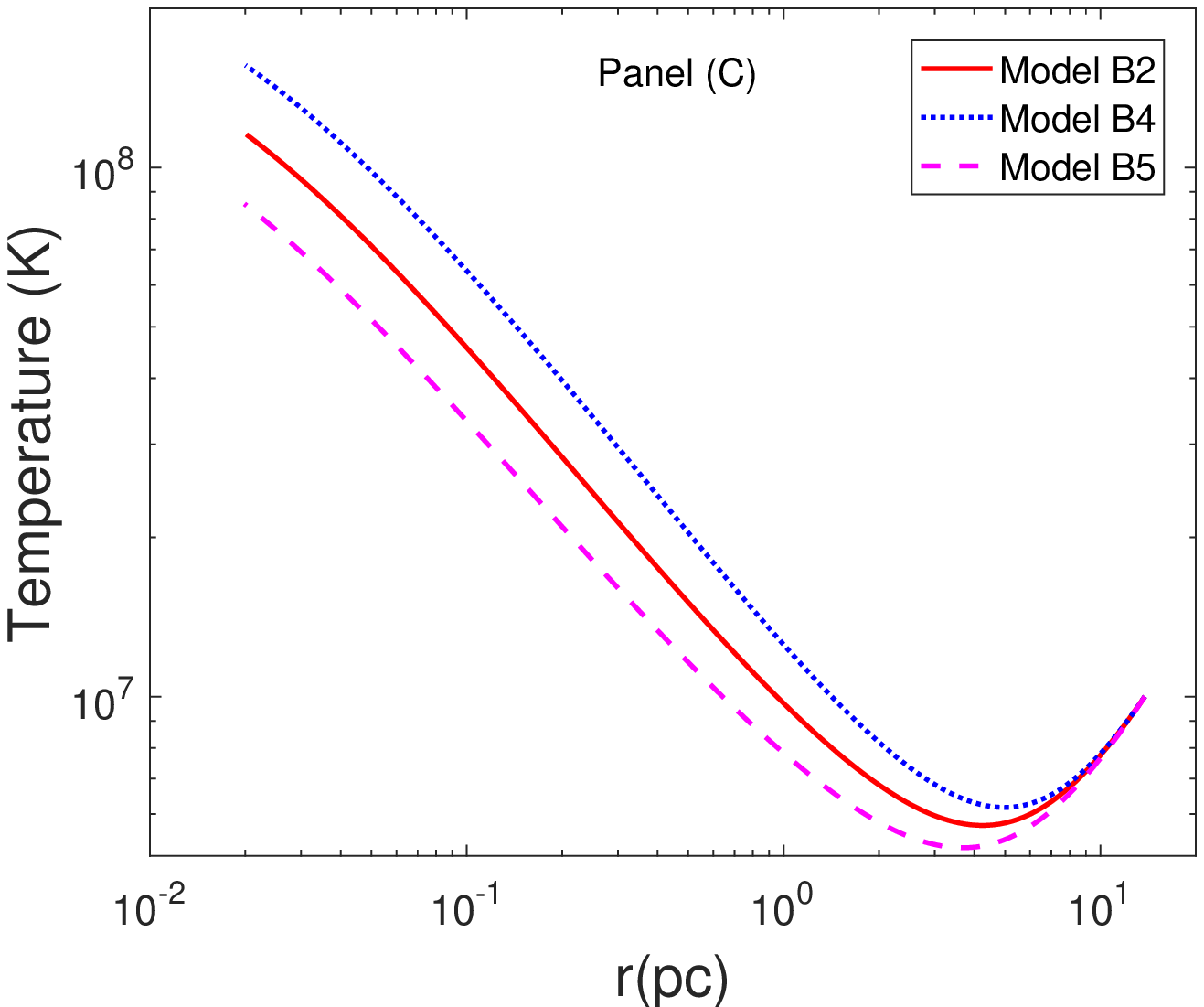}}}
\scalebox{0.5}[0.5]{\rotatebox{0}{\includegraphics{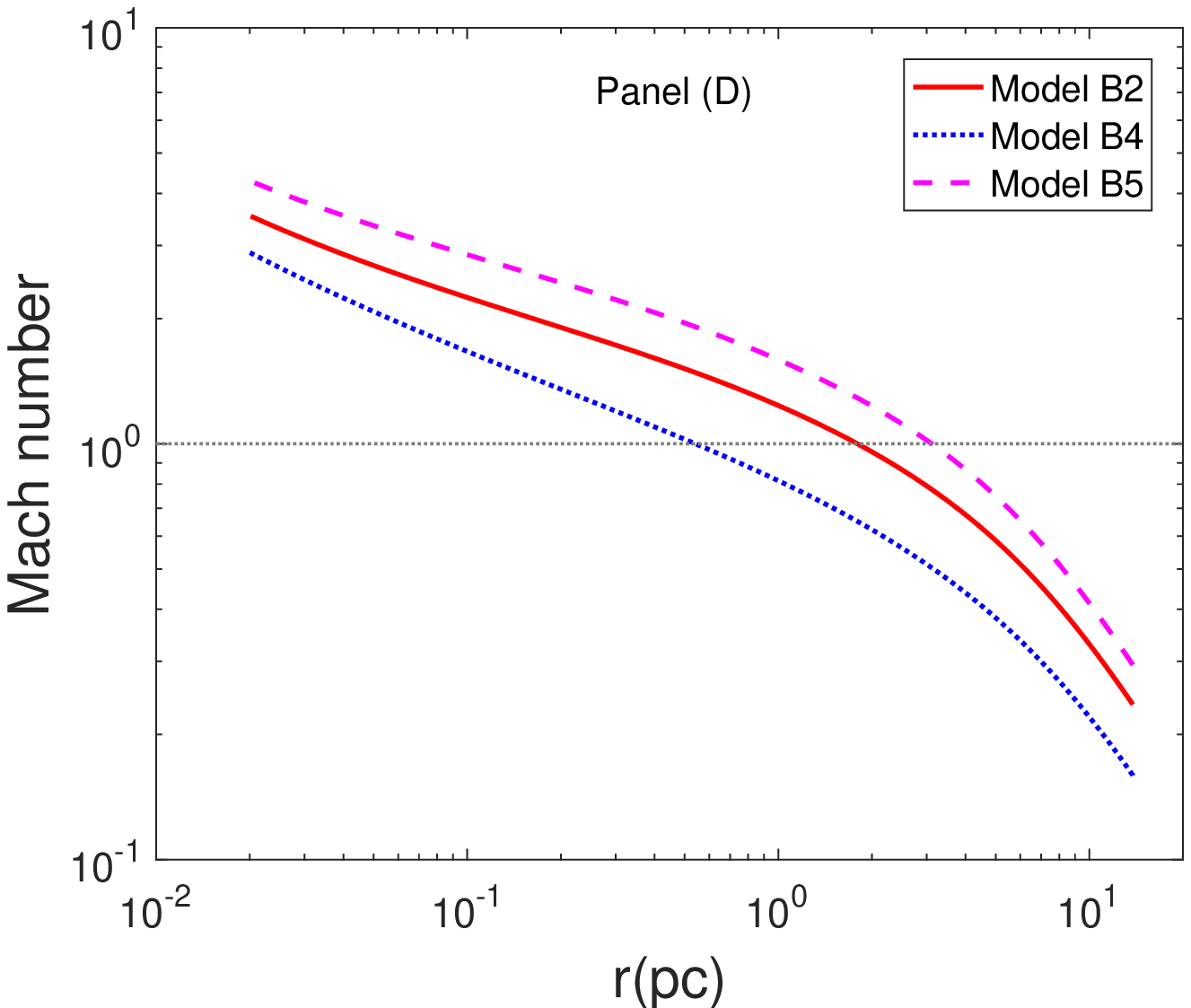}}}
\ \centering \caption{Radial dependent of velocity, temperature, density, and \textit{Mach} number in models B2, B4, and B5. }\label{fig 7}
\end{figure*}

\begin{figure*}
\scalebox{0.38}[0.38]{\rotatebox{0}{\includegraphics{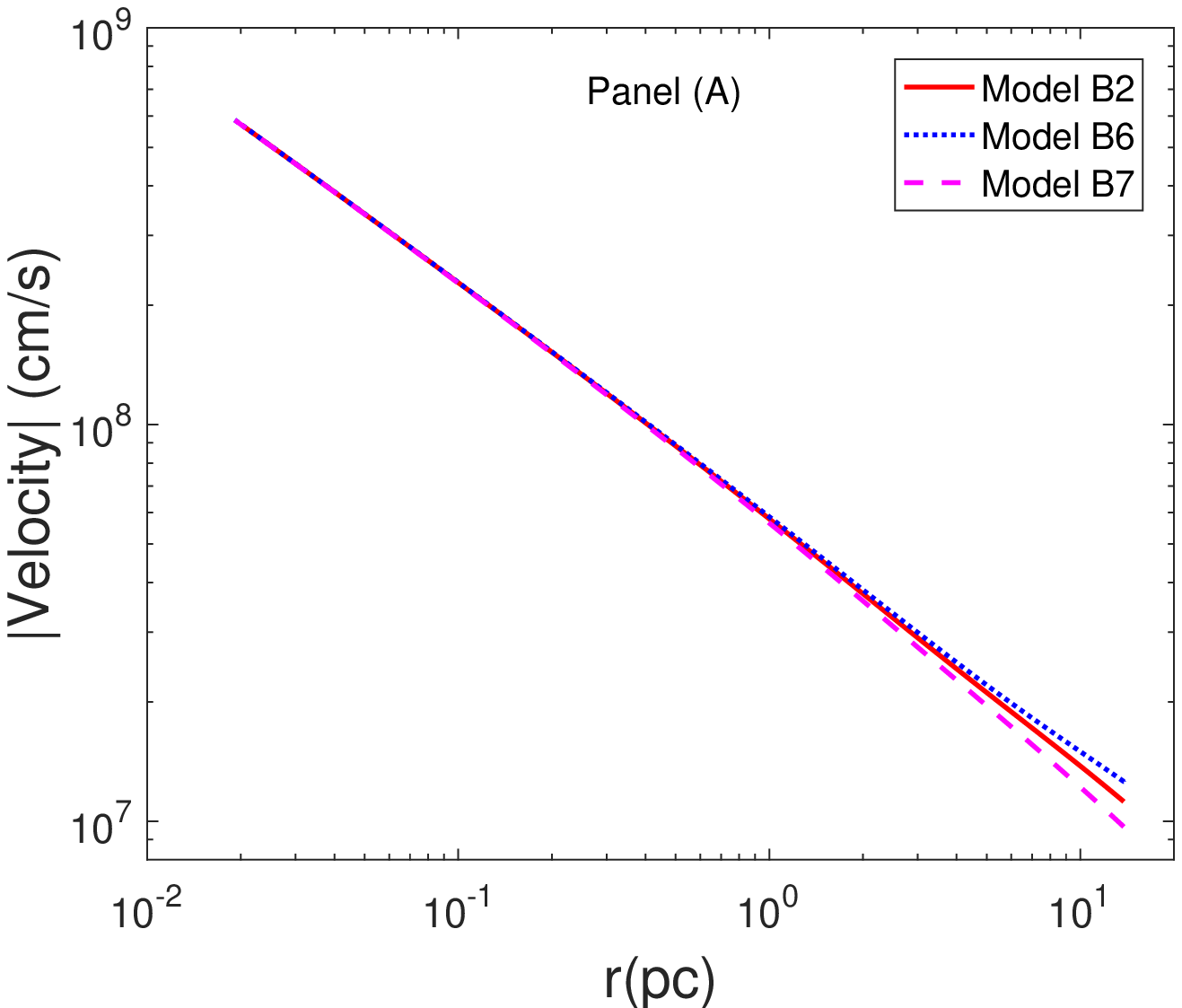}}}
\scalebox{0.38}[0.38]{\rotatebox{0}{\includegraphics{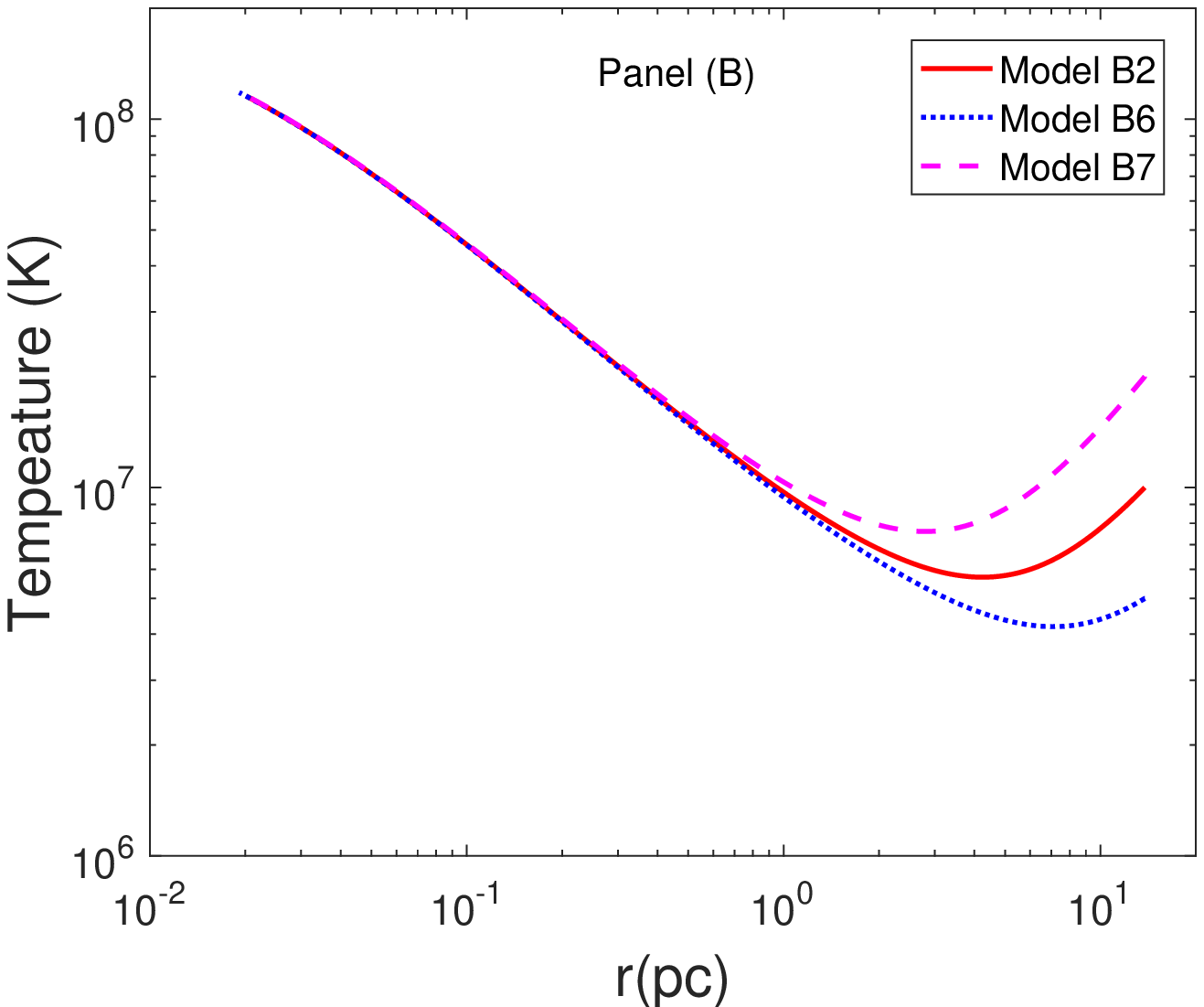}}}
\scalebox{0.38}[0.38]{\rotatebox{0}{\includegraphics{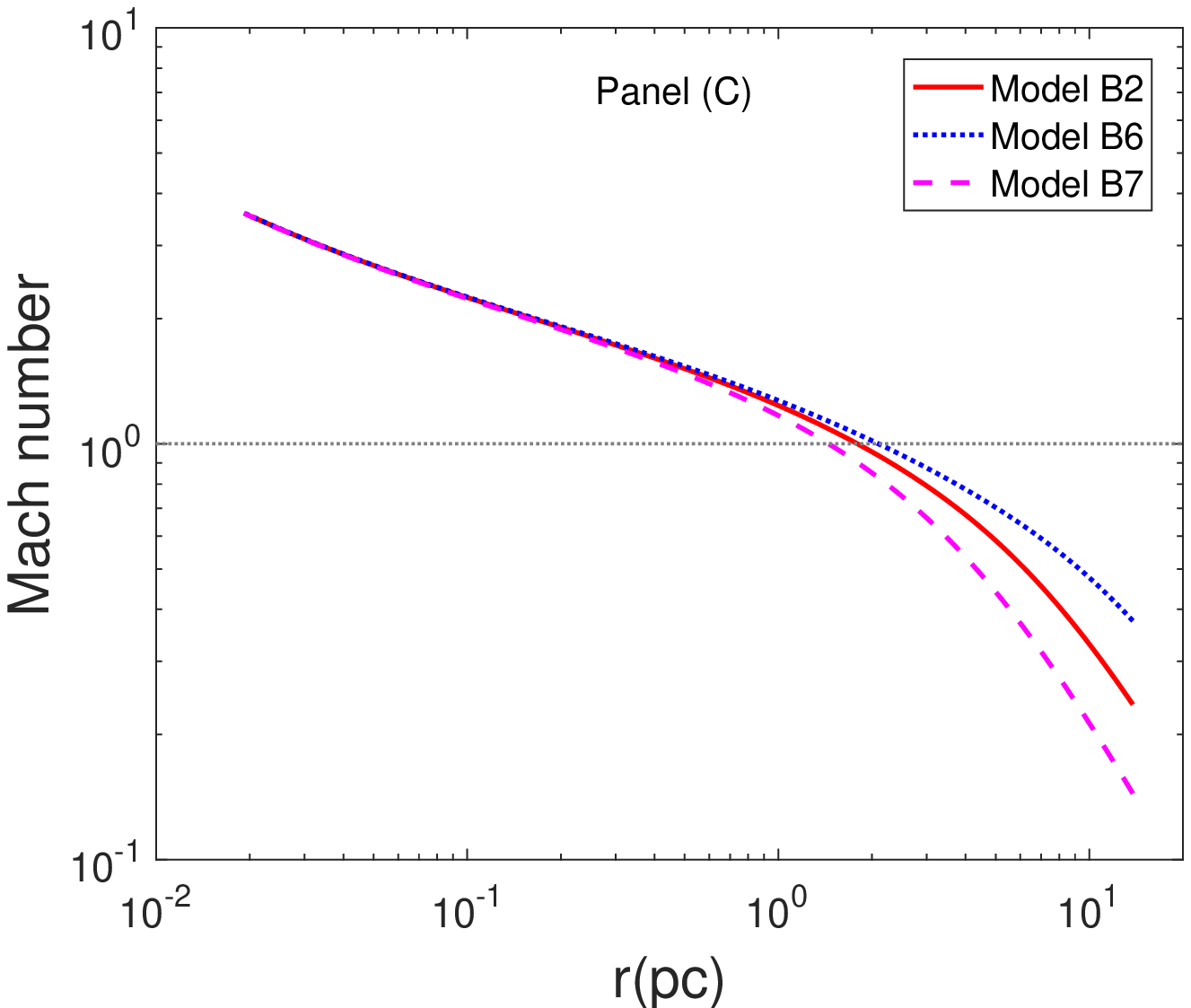}}}
\ \centering \caption{Radial dependent of velocity, temperature and \textit{Mach} number in models B2, B6, and B7. }\label{fig 8}
\end{figure*}
\begin{figure*}
\scalebox{0.38}[0.38]{\rotatebox{0}{\includegraphics{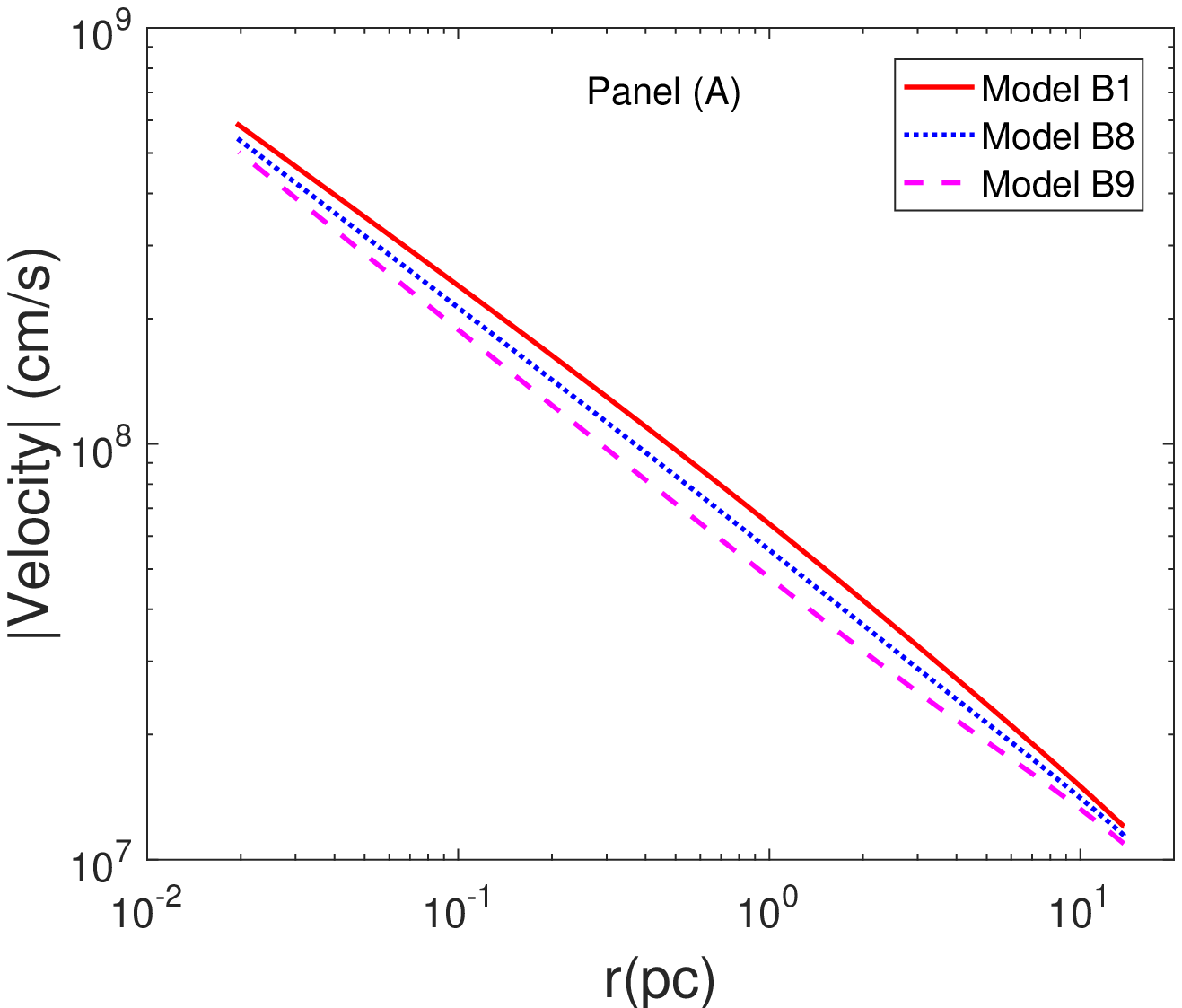}}}
\scalebox{0.38}[0.38]{\rotatebox{0}{\includegraphics{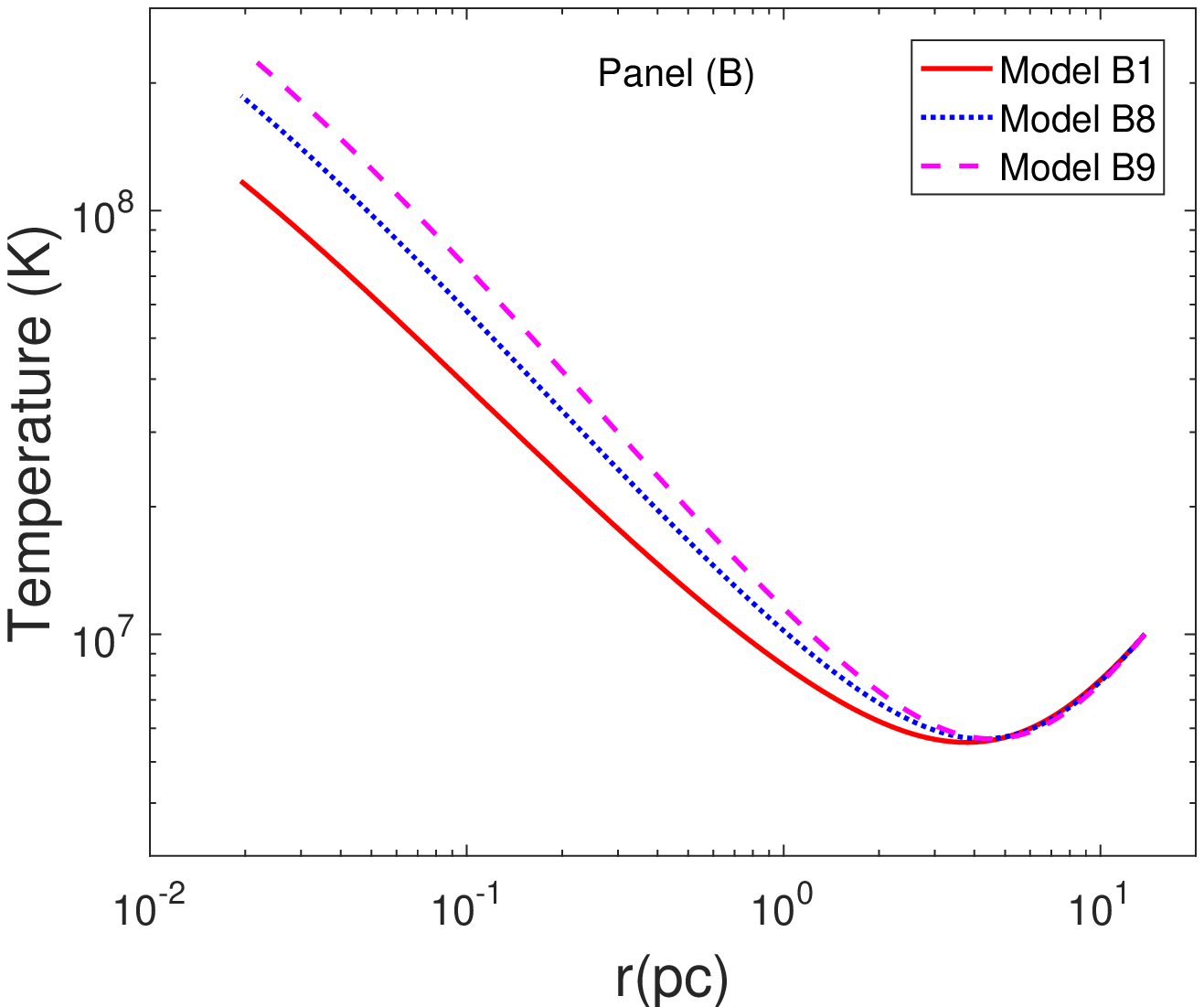}}}
\scalebox{0.38}[0.38]{\rotatebox{0}{\includegraphics{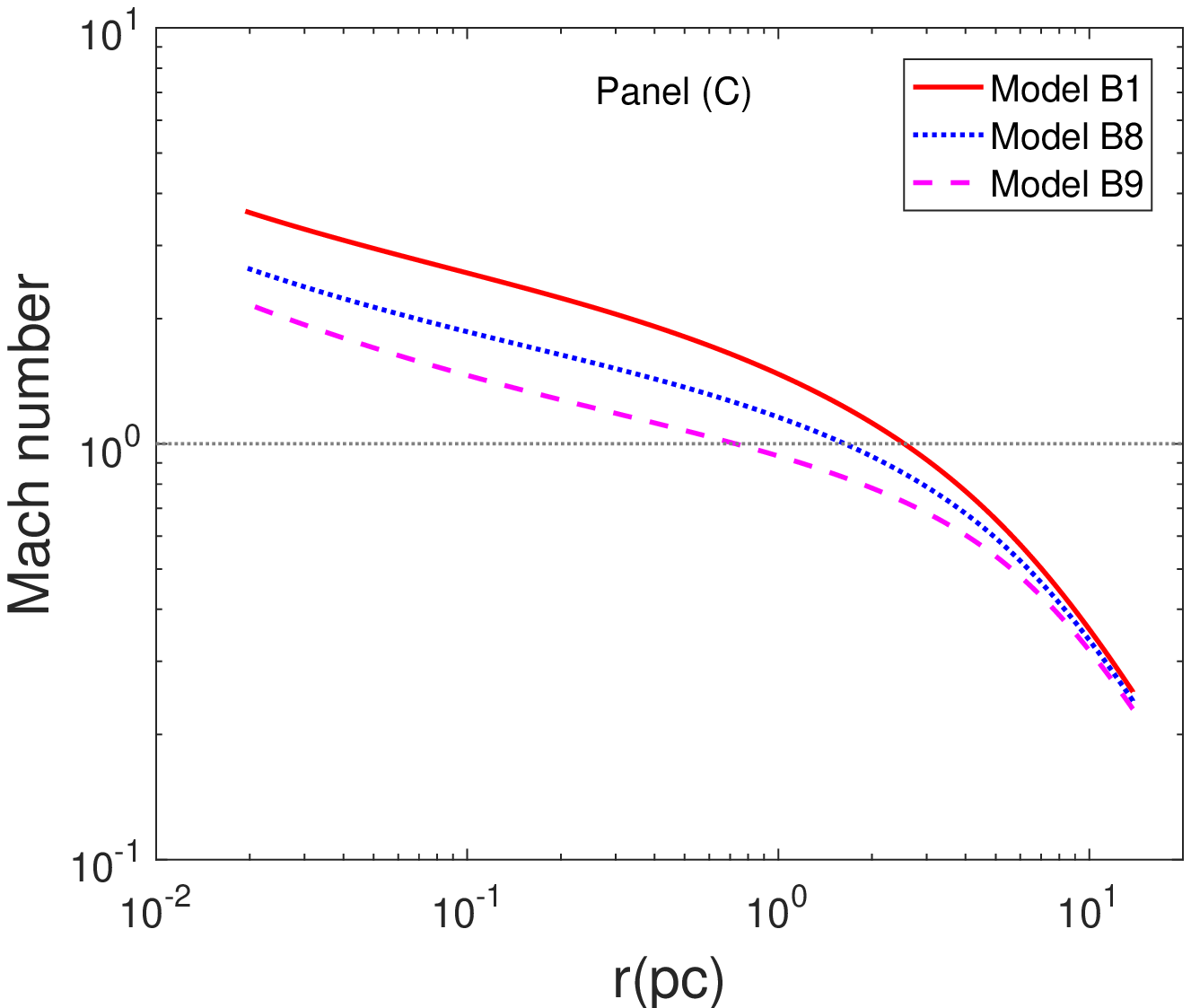}}}
\ \centering \caption{Radial dependent of velocity, temperature and \textit{Mach} number in models B1, B8, and B9. }\label{fig 9}
\end{figure*}

\begin{figure*}
\scalebox{0.55}[0.55]{\rotatebox{0}{\includegraphics{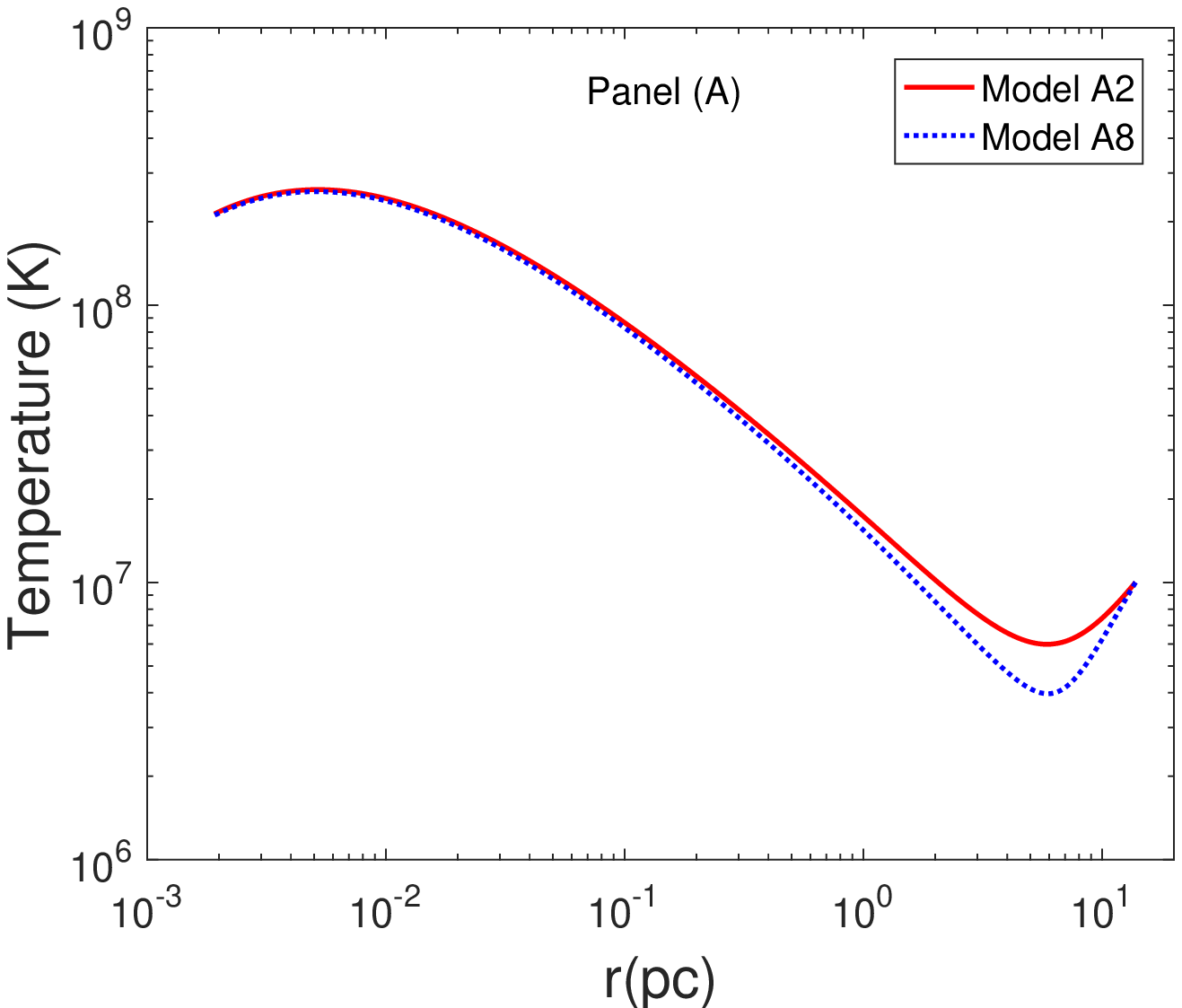}}}
\scalebox{0.55}[0.55]{\rotatebox{0}{\includegraphics{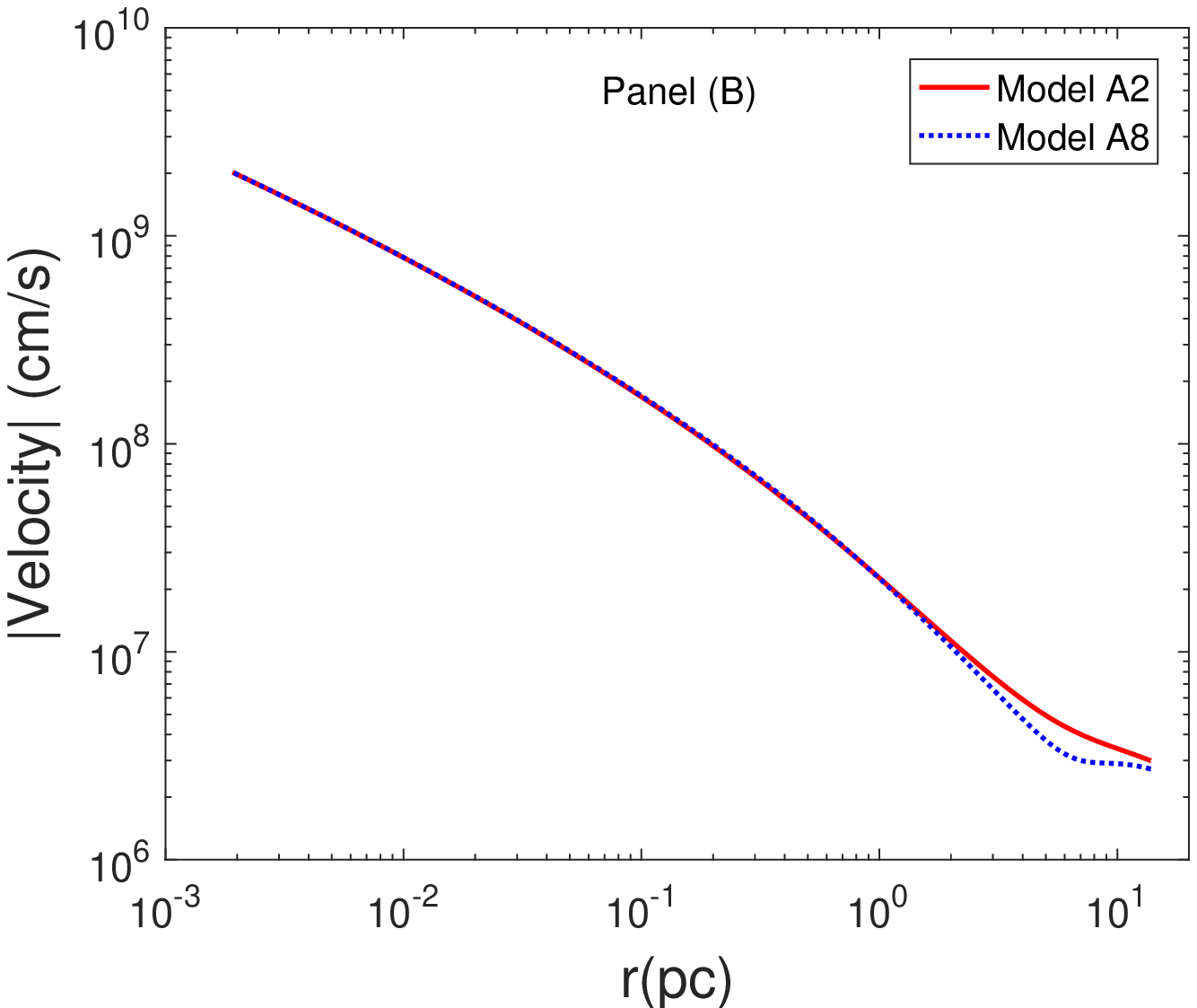}}}
\ \centering \caption{The comparison between the solutions of A2 and A8. }\label{fig 10}
\end{figure*}

In addition, in order to study the effects of $T(r_{\rm out})$ and $T_{\rm C}$ in the case with high accretion rate, we also compare models with different boundary temperature and different Compton temperature in Figures 8 and 9, respectively. The results shown in Figure 8 do not have significant differences from that shown in Figure 3. According to Figure 9, the radial dependence of radial velocity is similar in models B1, B8 and B9. This implies that different Compton temperature does not obviously change dynamic properties of models B1, B8 and B8. At large radii ($>4$ pc), their gas temperature is almost same. At the radii of $r<4$ pc, when Compton temperature is higher the gas temperature increases inwards faster. Comparing models B1, B8 and B9 with models A1, A6 and A7, we find that when the Compton temperature is higher, the difference of radial velocity at the radii of $>$1 pc is more obvious in the models with lower accretion rate (i.e. models A1, A6 and A7).

\subsection{Effect of the bulge stellar potential on accretion flows}
Bulge stellar potential becomes significant when the radius is beyond 1 pc (Yang \& Bu 2018b). At the radii of $>10$ pc, the bulge stellar gravity may be stronger than the black hole gravity. Here, we focus on the effect of the bulge stellar potential, especially at large radii. Figure 10 compares the solutions of models A2 and A8. Model A2 includes the bulge stellar potential while model A8 does not include it. As shown in Figure 10, the bulge stellar potential makes radial velocity increasing inwards slightly faster at the radii of $>2$ pc. At the radii of $<2$ pc where the stellar potential becomes weak, the radial dependence of radial velocity is almost the same as shown in models A2 and A8. Because the radial velocity is higher at the radii of $>2$ pc for model A2, the gas density decreases, and then radiative cooling becomes weaker. Therefore, the gas temperature in model A2 decreases slower than that in model A8 at large radii.

\subsection{Thermal instability}
Thermal instability of irradiated accretion flows at parsec-scale has been discussed in previous works (e.g. Krolik \& London 1983; Moscibrodzka \& Proga 2013). In Moscibrodzka \& Proga (2013), Compton temperature was set to be 2.9$\times 10^{7}$ K. However, the Compton temperature of LLAGNs radiation is higher. Here, we mainly analyze the effect of Compton temperature ($T_{\rm C}$) on the thermal instability.

When strong thermal instability occurs, it prevents us from obtaining a solution of equations (11) and (12). This is because that when thermally instability takes place, the accretion flows are not steady and evolve with time. Numerical simulations have identified that the gases become two phases, i.e. hot gases and cool gases, and the hot gases may become outflows (e.g. Moscibrodzka \& Proga 2013). The study of the unstable solutions is beyond the scope of this paper. Here, we follow Ostriker et al. (1976) and Krolik \& London (1983) to search the boundary between stability and instability. Krolik \& London (1983) mainly studied the effect of $T(r_{\rm out})$ on the boundary of stability and pointed out that a higher temperature ($T(r_{\rm out})$) helps to thermally stabilize the accretion flow. In order to study the effect of $T_{\rm C}$ on thermal instability, we set $T(r_{\rm out})$ to be $10^7$ K and then take $\dot{M}$, $L_{\rm X}$ and $T_{\rm C}$ as a parameter space. For the given $T_{\rm C}$, we change $\dot{M}$ and $L_{\rm X}$ and solve equations (11) and (12) until a stable-state solution can not be obtained. Figure 11 plots the stability boundary as three lines. These three lines correspond to different values of $T_{\rm C}$, respectively. In the lower right domain of the lines, we can always obtain a solution. However, the flow is thermally unstable in the higher left domain of the lines. As shown in figure 11, a higher $T_{\rm C}$ makes the accretion flow thermally unstable and a lower $\dot{M}$ also makes the accretion flow thermally unstable for a fixed $T_{\rm C}$.

For LLAGNs, their luminosity is often considered to be below 2\% $L_{\rm Edd}$, while the accretion rate may be beyond 2\% $\dot{M}_{\rm Edd}$ at the parsec scale. As shown in Figure 11, when the luminosity is less than 2\% $L_{\rm Edd}$, thermal instability may happen around LLAGNs at the sub-parsec or parsec scale. Moscibrodzka \& Proga (2013) have implemented numerical simulations and found the thermal instability when luminosity is less than 2\% $L_{\rm Edd}$. However, they adopted a lower $T_{\rm C}$. The temperature is suitable for a quasar.

We further analyze the reason for thermal instability. A linear analysis of thermal instability was formulated by Field (1965) and restated by (Moscibrodzka \& Proga 2013). In this paper, we follow the method used by Moscibrodzka \& Proga (2013) to analyze the thermal instability and explain why a higher ${T_{\rm{C}}}$ makes the parsec-scale accretion flow thermally unstable. We briefly introduce their theory here. When a small perturbation by $e^{\zeta t+ikx}$ is exerted on fluid equations, the dispersion relation is obtained as follows:
\begin{equation}
{\zeta ^3} + {N_v}{\zeta ^2} + {k^2}c_s^2\zeta  + {N_p}{k^2}c_s^2 = 0,
\end{equation}
where $k$ is the perturbation wave number. Two growth rate functions ${N_{\rm{v}}}$ and ${N_{\rm{p}}}$ are defined as (Moscibrodzka \& Proga 2013),
\begin{equation}
{N_{\rm{p}}} \equiv \frac{1}{{{c_{\rm{p}}}}}{(\frac{{\partial {\cal L}}}{{\partial T}})_p}
\end{equation}
and
\begin{equation}
{N_{\rm{v}}} \equiv \frac{1}{{{c_{\rm{v}}}}}{(\frac{{\partial {\cal L}}}{{\partial T}})_\rho },
\end{equation}
where ${c_{\rm{p}}}$ and ${c_{\rm{v}}}$ are specific heats when pressure or volume are constant, respectively, and ${\cal L} = -{n^2}S/\rho $ is the net radiative cooling rate per unit of mass. We further calculate the growth timescales ${\tau _{{\rm{growth}}}} = \frac{1}{\zeta }$ for different modes as :
\begin{equation}
{\tau _{{\rm{TI}}}} =  - \frac{1}{{{N_{\rm{p}}}}}, {\tau _{\rm{v}}} =  - \frac{1}{{{N_{\rm{v}}}}},\text{ and }{\tau _{{\rm{ac}}}} =  - \frac{2}{{\left( {{N_{\rm{v}}} - {N_{\rm{p}}}}, \right)}}
\end{equation}
where ${\tau _{{\rm{TI}}}}$ is the growth timescale of short wavelength, isobaric condensations, ${\tau _{\rm{v}}}$ is the growth timescale of the long wavelength, isochoric perturbations, and ${\tau _{{\rm{ac}}}}$ is the growth timescales of the short wavelength, isentropic sound waves, respectively. When the growth timescale is negative, the small perturbation is damped. When the growth timescale is positive, thermally instability grows in the falling gas. However, when the ratio of the accretion timescale (${\tau _{{\rm{acc}}}} = r/v$) to the growth timescale (${\tau _{{\rm{growth}}}}$) is relatively small, the perturbation grows so slow that the gas leaves the unstable zone before the instability obviously affects the accretion flow (Krolik \& London 1983). In this case, the flow is ``marginally'' stable and we can also obtain a solution. This case makes it necessary to compare ${\tau _{{\rm{acc}}}}$ and ${\tau _{{\rm{growth}}}}$. Therefore, we plot the radial dependence of ${\tau _{{\rm{acc}}}}/{\tau _{{\rm{growth}}}}$ in Figure 12. When the ${\tau _{{\rm{acc}}}}/{\tau _{{\rm{growth}}}}$ value exceeds a critical value, we can not obtain a solution because of the strong thermal instability. This critical value depends on the integral step size but should have an order of magnitude of 1.

\begin{figure}
\scalebox{0.6}[0.6]{\rotatebox{0}{\includegraphics{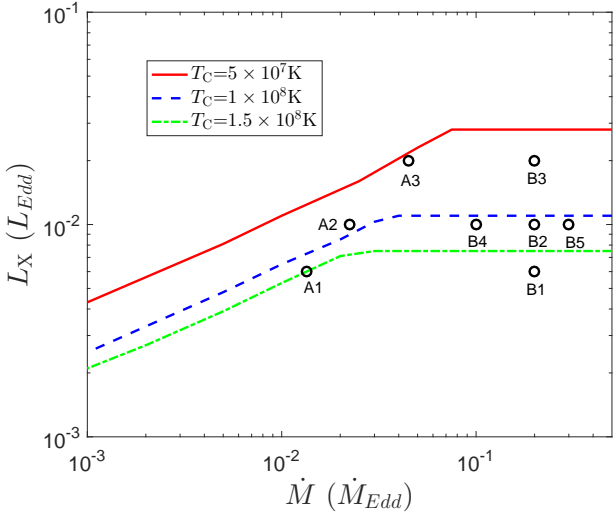}}}
\ \centering \caption{Boundaries between stability and instability in the ${L_{X}} - \dot M$ plot. The red solid, blue dashed and green dash-dotted lines are the boundaries of instability for three Compton temperature $5 \times {10^7}$K, $1 \times {10^8}$K and $1.5 \times {10^8}$K, respectively. Beneath those boundaries we can always calculate a solution. Above but close to those boundaries it is possible to find a stable solution sometimes. Far above the boundaries the gas is totally unstable. The models marked by circles are at least ``marginally'' stable.}\label{fig 11}
\end{figure}
\begin{figure}
\scalebox{0.6}[0.6]{\rotatebox{0}{\includegraphics{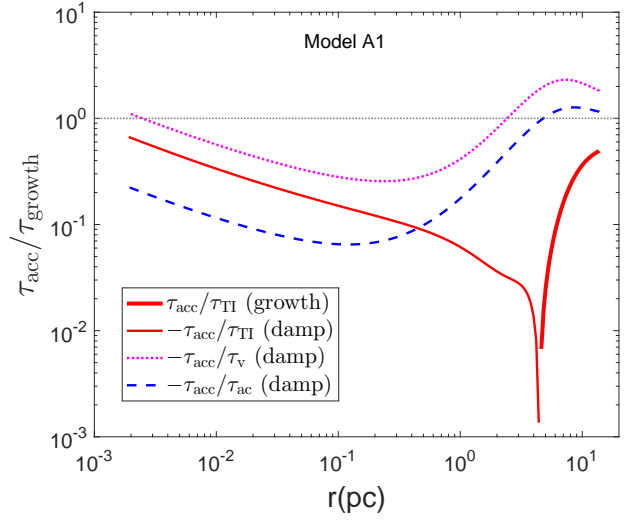}}}
\scalebox{0.6}[0.6]{\rotatebox{0}{\includegraphics{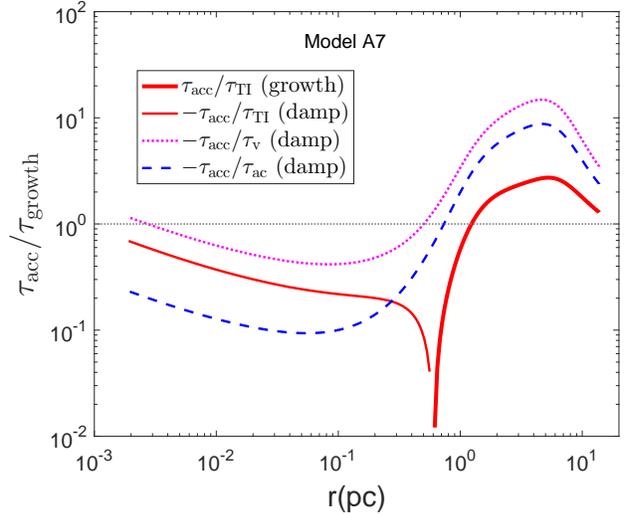}}}
\ \centering \caption{The radial dependence of ${\tau _{{\rm{acc}}}}/{\tau _{{\rm{growth}}}}$ for models A1
(top) and A7 (bottom).}\label{fig 12}
\end{figure}

In Figure 12, we find that both of the long-wavelength, isochoric perturbations, and the short-wavelength, isentropic sound waves are damped at all radii, while the short wavelength, isobaric condensations grow at large radii and cause thermal instability here. This is similar to the results in Moscibrodzka \& Proga (2013). In their results, only the short wavelength, isobaric condensations grow at larger radii.

Comparing models A1 and A7, ${\tau _{{\rm{acc}}}}/{\tau _{{\rm{TI}}}}$ is larger in model A7 than in model A1 at large radii. In model A7, Compton temperature is higher. This indicates that the ${\tau _{{\rm{acc}}}}/{\tau _{{\rm{TI}}}}$ values in the models with a higher Compton temperature are larger than that in the models with a lower Compton temperature. As a result, when a higher Compton temperature is adopted, the condensation mode of the gas can grow a greater amount in the unstable region, which means that the models with a higher Compton temperature are more unstable. Since a lower luminosity helps to stabilize the flow (Ostriker et al. 1976; Krolik \& London 1983; Moscibrodzka \& Proga 2013), the boundary of stability moves to the region with lower luminosity when a higher Compton temperature is adopted. Model A7 is located at the stability boundary in Figure 11. This implies that when the ${\tau _{{\rm{acc}}}}/{\tau _{{\rm{TI}}}}$ value becomes larger than that of model A7, the accretion flow becomes thermally unstable. According to Figure 12, the critical value of ${\tau _{{\rm{acc}}}}/{\tau _{{\rm{TI}}}}$ is 3. In other words, when ${\tau _{{\rm{acc}}}}/{\tau _{{\rm{TI}}}} > 3$, the thermal instability prevents us from calculating a stable solution.

\subsection{The dependence of accretion rates on the gas density and temperature}

An important application of our models is to estimate the mass accretion rates when the gas density and temperature at the parsec scale are given. In most cases, the classical Bondi model (Bondi, 1952) is used to predict the accretion rates with an analytical formula
\begin{equation}
{\dot M_{{\rm{Bondi}}}} = \pi {r_{\rm{B}}}^2{\rho _\infty }{c_\infty },
\end{equation}
where ${r_{\rm{B}}} = G{M_{{\rm{BH}}}}/{c_\infty }^2$ is the Bondi radius, ${\rho _\infty }$ and ${c_\infty }$ are the density and acoustic velocity of the gas at infinity. The Bondi model predicts $\dot{M}_{\rm Bondi}\propto \rho_{\infty}T^{-1.5}_{\infty}$, where $T_{\infty}$ the gas temperature at infinity. When we calculate the Bondi accretion rate, the gas density ($\rho(r_{\rm out})$) and temperature  ($T(r_{\rm out})$) at the outer boundary are considered as those at infinity. The classical Bondi model may be too simple and lacks many necessary details for realistic accreting processes, such as radiative heating and cooling. Our models have modified the classical Bondi model. Figure 13 shows the dependence of accretion rates on $\rho(r_{\rm out})$) and $T(r_{\rm out})$. In Figure 13, red lines mean the accretion rates ($\dot{M}_{\rm A}$) predicted by A-type models while blue lines mean ${{\dot M}_{{\rm{A}}}}/{{\dot M}_{{\rm{Bondi}}}}$.

As shown in the top panel of Figure 13, with the increase of $\rho(r_{\rm out})$, the accretion rates (red lines) predicted by A-type models increase faster than the changing trend predicted by the Bondi model. The bottom panel shows that, with the increase of $T(r_{\rm out})$, the accretion rates (red lines) of A-type models decrease slower than the changing trend predicted by the Bondi model. Compared to A-type models, the Bondi model always underestimates the accretion rates, as shown by blue lines. According to the results given in section 3.1, there is always significant net radiative cooling at large radii due to that bremsstrahlung cooling is dominant, which provides an extra inwards pressure force here. Therefore, the gas velocity in our models is higher than the gas velocity in the Bondi model. This causes the accretion rates to be underestimated in the Bondi model. When the Compton temperature decreases, the net radiative cooling at large radii is slightly strengthened. This is helpful to increase the accretion rates in A-type models.

\begin{figure}
\scalebox{0.6}[0.6]{\rotatebox{0}{\includegraphics{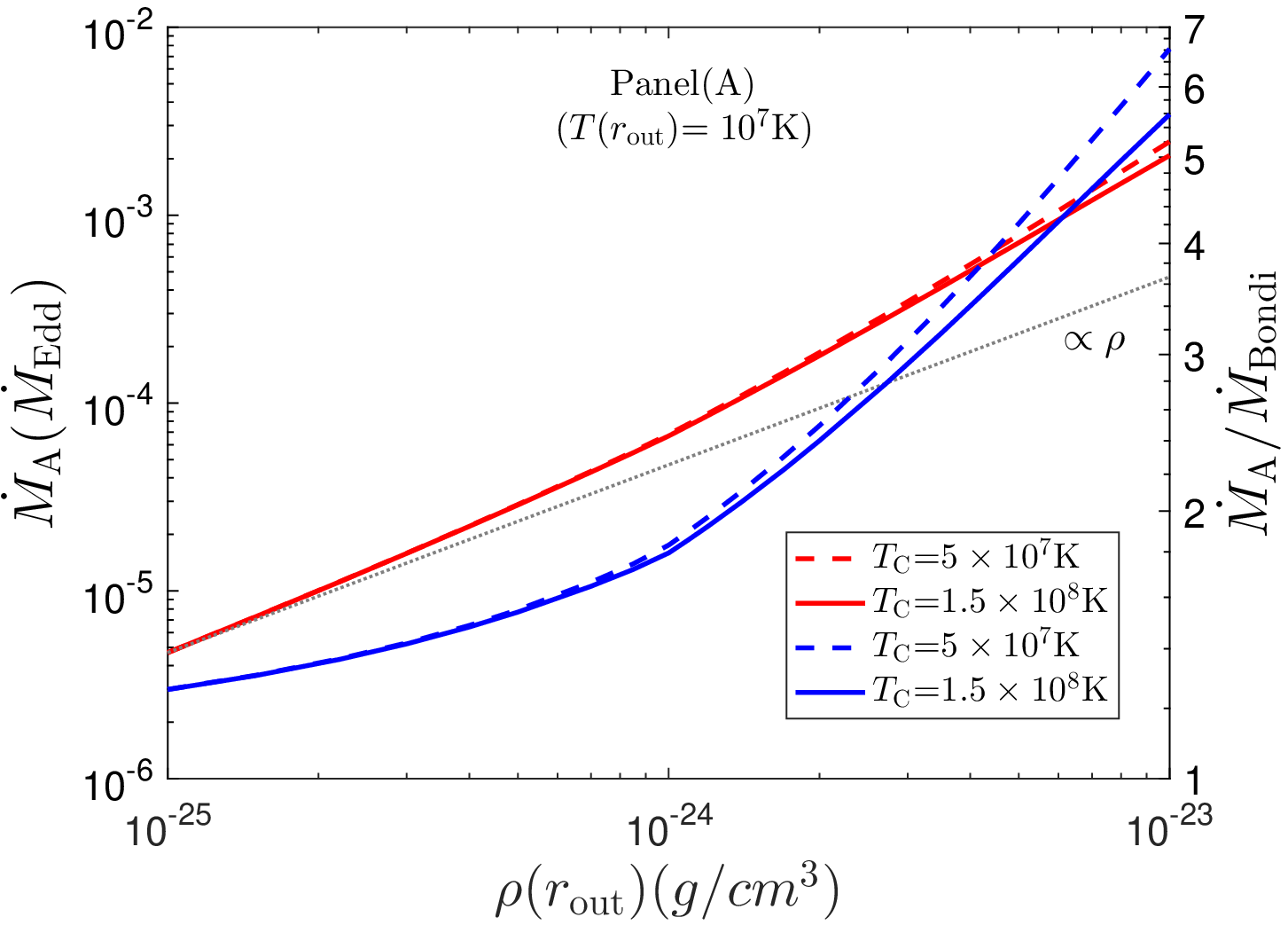}}}
\scalebox{0.6}[0.6]{\rotatebox{0}{\includegraphics{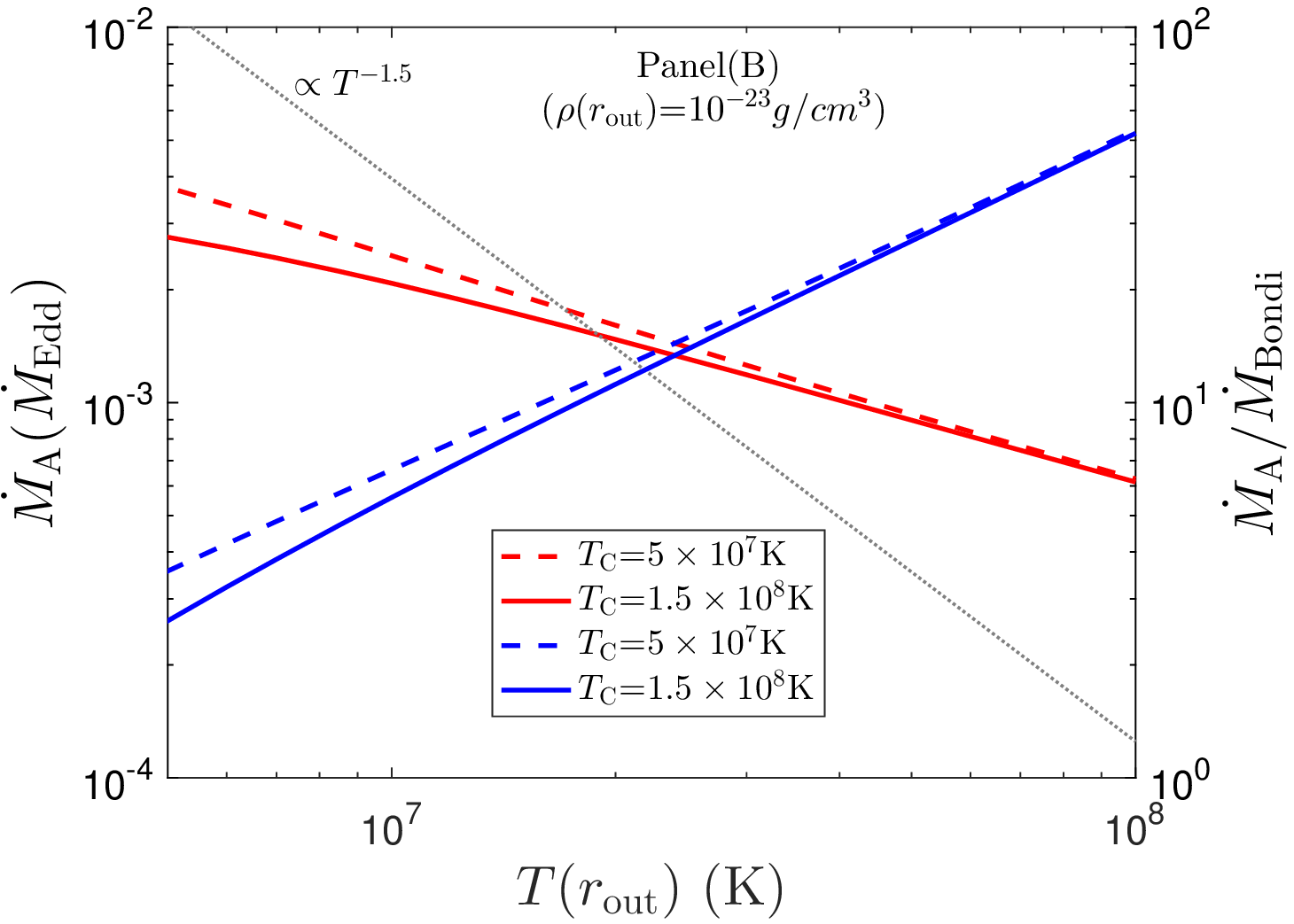}}}
\ \centering \caption{The dependence of accretion rates on the gas density (top) and temperature (bottom) at the outer boundary. In the top panel, we set the gas temperature at the outer boundary to be $10^7$ K. In the bottom panel, we set the gas density at the outer boundary to be $10^{-23}$ g$\cdot$cm$^{-3}$. Red lines mean the accretion rates ($\dot{M}_{\rm A}$) predicted by A-type models while blue lines mean ${{\dot M}_{{\rm{A}}}}/{{\dot M}_{{\rm{Bondi}}}}$. Black dotted lines mean the changing trend of the accretion rates predicted by the Bondi model.}\label{fig 12}
\end{figure}

\subsection{Comparison to observations}
The hot gas density and temperature around LLAGNs were measured at X-ray bands. This provides a chance for comparison to observations. Pellegrini (2005) has collected observational data from references. Among those data, the observed gas density and temperature of four LLAGNs (NGC 221, NGC 821, NGC 1553, and NGC 4438) refer to the radii much large than their Bondi radius. The radii are comparable with the outer boundary in A-type models. For the other LLAGNs, their gas density and temperature almost refer to their Bondi radius. When the Bondi radius is set to be the outer boundary in A-type models, the radiative cooling region at large radii is not in the range of our calculation. In this case, the accretion rate predicted by A-type models is comparable to the Bondi model.  Therefore, we mainly focus on the four LLAGNs here. Table 2 lists the observed properties of the four LLAGNs and the $L_{{\rm{A}}}/L_{{\rm{obs}}}$, where $L_{\rm A}$ is the bolometric luminosity predicted by A-type models while $L_{\rm obs}$ is the observed X-ray luminosity given by Pellegrini (2005). For LLAGNs, the X-ray luminosity is almost dominant in the bolometric luminosity. For NGC 221, NGC 1553, and NGC 4438, the ratio of the predicted \textbf{X-ray} luminosity to the observed X-ray luminosity is within one order of magnitude. This result is acceptable to some degree. It is noted that the net accretion rate predicted by A-type models depends on $r_{\rm cir}$, and then the luminosity  predicted by A-type models also depends $r_{\rm cir}$.

For the sake of comparison, we also give the $L_{{\rm{Bondi}}}/L_{{\rm{obs}}}$ in table 2, where $L_{{\rm{Bondi}}}$ is the bolometric luminosity predicted by the Bondi model. When the Bondi model is used to predict a luminosity, we still adopt the radiative efficiency given by equation (10). However, the accretion rate given by the Bondi model is taken as the net accretion rate in Equation (10). In this case, although the accretion rate predicted by Bondi model is relatively lower, as shown in Figure 13, the estimated luminosity by the Bondi model is higher than that by A-type models.

\begin{table}
\begin{center}

\caption[]{Comparison to observations}
\scriptsize
\begin{tabular}{ccccccccccc}
\hline\noalign{\smallskip} \hline\noalign{\smallskip}

 Objects&  $M_{\rm BH}$ & $\rho_{\rm obs}$  &  $T_{\rm obs}$ & $ \frac{L_{\rm A}}{L_{\rm obs}}$ &$\frac{{{L_{{\rm{Bondi}}}}}}{{{L_{{\rm{obs}}}}}}$\\
    &   ($10^8 M_{\odot}$)   & (${10^{ - 24}}$g$\cdot$cm$^3$) & (${10^{6}}$K) & &  \\
(1) & (2)     & (3) &  (4)      &     (5)   & (6)\\

\hline\noalign{\smallskip}
NGC221 &0.025 &0.13&4.3&0.31&1.1  \\
NGC821 &0.37  &0.01&5.3&$>$0.024&$>$0.073  \\
NGC1553&1.6   &0.06&5.9&7.4&24  \\
NGC4438&0.5   &0.99&6.7&7.5&17  \\
\hline\noalign{\smallskip}
\hline\noalign{\smallskip}
\end{tabular}
\end{center}

\begin{list}{}
\item\scriptsize{\textit{Note}. Column (1): the name of galaxies; Column (2): the BH mass; Columns (3) and (4): the observed gas density ($\rho_{\rm obs}$) and temperature ($T_{\rm obs}$), respectively; Column (5): the ratio of the predicted luminosity by A-type models (${L_{\rm A}}$) to the observed X-ray luminosity ($L_{\rm{obs}}$). Column (6): the ratio of the predicted luminosity by Bondi models (${L_{\rm Bondi}}$) to the observed X-ray luminosity ($L_{\rm{obs}}$).}
\end{list}
\label{table2}
\end{table}
\section{SUMMARY AND DISCUSSION}
The properties of gases at the parsec-scale can significantly influence the activity of LLAGNs. Here, we analytically study the dynamical and thermal properties of the parsec-scale gases when they are accreted on LLAGNs, which have higher Compton temperatures than quasars. The parsec-scale gases are irradiated by LLAGNs. Therefore, we take into account Compton heating/cooling and photoionization heating by the X-ray radiation from LLAGNs. We also consider radiation cooling, such as the bremsstrahlung cooling and the recombination and line cooling. Bulge stellar potential is also taken into account.

In this paper, we study the effects of a set of parameters (i.e. luminosity, mass accretion rate, the temperature at the outer boundary, and Compton temperature) on thermal and dynamical properties of the parsec-scale gases and analyze thermal instability. Using the analytical method described in section 2.3, we obtain a serial of steady solutions. Our main results are summarized in the following.

(1) When the radiative heating/cooling is included in models, the model solutions obviously deviate from Bondi solution in thermal and dynamical properties. At different radii, the thermal properties of gases are different. At large radii (e.g. $>$ 4 pc), thermodynamic processes are dominated by bremsstrahlung cooling. At medium radii (e.g. 0.01--4 pc ), compressional heating is dominant. At small radii (e.g. $<$ 0.01 pc), Compton cooling is important. For dominated thermodynamic processes, the spatial domain in which they work and their strength is different for different model parameters. As a result, different model parameters make the radial dependence of radial velocity and gas temperature different. We have discussed our results in detail, in sections 3.1 and 3.2.

(2) We give the boundary between thermal stability and instability, as shown in Figure 11. Thermal stability is attributed to the growth of the short wavelength, isobaric condensations at large radii. ${\tau _{{\rm{acc}}}}/{\tau _{{\rm{TI}}}}=3$ could be a critical value of thermal stability. When ${\tau _{{\rm{acc}}}}/{\tau _{{\rm{TI}}}}>3$, thermal instability could take place. We find that a higher Compton temperature makes the ${\tau _{{\rm{acc}}}}/{\tau _{{\rm{TI}}}}$ value higher, which means that thermal instability is stronger when Compton temperature is higher. As a result, Compton temperature significantly influences the boundary between thermal stability and instability. A higher Compton temperature easily makes the falling gases thermally unstable.

(3) Compared to our models, the Bondi model underestimates the accretion rate. When radiative cooling and heating are included, the gases cool at the large radii due to bremsstrahlung cooling. This is helpful to increase the inward velocity of falling gas, compared to the Bondi model. We have used our models to estimate the luminosity of NGC 221, NGC 1553, and NGC 4438. We find that the ratio of the estimated luminosity to the observed luminosity is within one order of magnitude for three sources.

Due to the inherent limitation of the spherically symmetric and time-independent models, we cannot study the solutions of the thermally unstable gas and the influence of the winds from the inner region. Thermal instability can cause gases to become two phases (i.e. hot gases and cool gases) and the hot gases may become outflows. This effect may reduce the accretion rates. When Compton temperature is different, it is necessary to numerically simulate the properties of thermally unstable gases in the future.

\section{ACKNOWLEDGMENTS}
This work is supported by the Natural Science Foundation of China (grant 11973018) and Chongqing Natural Science Foundation (grant cstc2019jcyj-msxmX0581). We thank the anonymous referee for the constructive suggests.

\section{DATA AVAILABILITY}
The data underlying this article will be shared on reasonable request to the corresponding author.
¡¤

\begin{appendix}
\section{DERIVATION OF EQUATIONS (11) AND (12)}

In the following, we describe the derivation of Equations (11) and (12). Equation (1) can be written as
\begin{equation}
\frac{{{\rm{d}}\ln v}}{{{\rm{d}}\ln r}} = -2 - \frac{{{\rm{d}}\ln \rho }}{{{\rm{d}}\ln r}}.
\end{equation}
According ${\cal M}^2 = \rho v^2/\gamma p$, we have
\begin{equation}
\frac{{{\rm{d}}\ln (\gamma {\mathcal M})}}{{{\rm{d}}\ln r}} = \frac{{{\rm{d}}\ln v}}{{{\rm{d}}\ln r}} - \frac{1}{2}\frac{{{\rm{d}}\ln p}}{{{\rm{d}}\ln r}} + \frac{1}{2}\frac{{{\rm{d}}\ln \rho }}{{{\rm{d}}\ln r}}.
\end{equation}
Using Equation (A1), the above equation is reduced to
\begin{equation}
\frac{{{\rm{d}}\ln (\gamma \mathcal M)}}{{{\rm{d}}\ln r}} =  - 2 - \frac{1}{2}\frac{{{\rm{d}}\ln p}}{{{\rm{d}}\ln r}} - \frac{1}{2}\frac{{{\rm{d}}\ln \rho }}{{{\rm{d}}\ln r}}.
\end{equation}

Equation (2) can be rewritten as
\begin{equation}
\frac{{{\rm{d}}\ln (\gamma {\mathcal M})}}{{{\rm{d}}\ln r}} = ( - \frac{1}{{\gamma {{\cal M}^2}}} - \frac{1}{2})\frac{{{\rm{d}}\ln p}}{{{\rm{d}}\ln r}} + \frac{1}{2}\frac{{{\rm{d}}\ln \rho }}{{{\rm{d}}\ln r}} - \frac{{r\rho }}{{\gamma {{\mathcal M}^2}p}}g.
\end{equation}
Combining Equations (A3) and (A4), we have
\begin{equation}
 - \frac{1}{{\gamma {{\cal M}^2}}}\frac{{{\rm{d}}\ln p}}{{{\rm{d}}\ln r}} + \frac{{{\rm{d}}\ln \rho }}{{{\rm{d}}\ln r}} - \frac{{r\rho }}{{\gamma {{\cal M}^2}p}}g + 2 = 0
\end{equation}
According to $p = (\gamma  - 1)e$, Equation (4) can be rewritten as
\begin{equation}
\frac{{pv}}{{r(\gamma  - 1)}}(\frac{{{\rm{d}}\ln p}}{{{\rm{d}}\ln r}} - \gamma \frac{{{\rm{d}}\ln \rho }}{{{\rm{d}}\ln r}}) = {n^2}S.
\end{equation}

We can solve $\frac{\rm{d} \rm{ln} p}{\rm{d} \ln r}$ and $\frac{\rm{d} \ln \rho}{\rm{d} \ln r}$ from Equations (A5) and (A6). $\frac{\rm{d} \rm{ln} p}{\rm{d} \rm {ln} r}$ and $\frac{\rm{d} \rm{ln} \rho}{\rm{d} \rm {ln} r}$ are given by
\begin{equation}
\frac{{{\rm{d}}\ln p}}{{{\rm{d}}\ln r}} =  - \frac{{2\gamma {{\mathcal M}^2}}}{{{{\mathcal M}^2} - 1}} + \frac{{\rho r}}{{({{\mathcal M}^2} - 1)p}}g + \frac{{{{\mathcal M}^2}r(\gamma  - 1)}}{{({{\mathcal M}^2} - 1)pv}}{n^2}S,
\end{equation}
and
\begin{equation}
\frac{{{\rm{d}}\ln \rho }}{{{\rm{d}}\ln r}} =  - \frac{{2{{\mathcal M}^2}}}{{{{\mathcal M}^2} - 1}} + \frac{{\rho r}}{{\gamma ({{\mathcal M}^2} - 1)p}}g + \frac{{r(\gamma  - 1)}}{{({{\mathcal M}^2} - 1)pv\gamma }}{n^2}S,
\end{equation}
respectively. According to the equation of state, we have
\begin{equation}
\begin{aligned}
\frac{{{\rm{d}}\ln T}}{{{\rm{d}}\ln r}} & = \frac{{{\rm{d}}\ln p}}{{{\rm{d}}\ln r}} - \frac{{{\rm{d}}\ln \rho }}{{{\rm{d}}\ln r}} \\
&  =  - \frac{{2{{\mathcal M}^2}(\gamma  - 1)}}{{{{\mathcal M}^2} - 1}} + \frac{{\rho r(\gamma  - 1)}}{{\gamma ({{\mathcal M}^2} - 1)p}}g + \frac{{r(\gamma  - 1)(\gamma {{\mathcal M}^2} - 1)}}{{({{\mathcal M}^2} - 1)pv\gamma }}{n^2}S.
\end{aligned}
\end{equation}
According to $\dot M = 4\pi {r^2}\rho v$ and Equation (3), Equations (A8) and (A9) are identical to Equations (11) and (12).
\end{appendix}

\end{document}